%% file: paper.tex
\documentclass[aps,prd,twocolumn,amsfonts,amssymb,amsmath,showpacs,superscriptaddress,letterpaper]{revtex4}

\usepackage{graphicx}
\usepackage{dcolumn}
\usepackage{color}

\newcommand{\BaBarYear}{07}
\newcommand{\BaBarNumber}{003}
\newcommand{\SLACPubNumber}{12668}
\newcommand{\BaBarType}     {CONF}  

\def\optbar#1{\vbox{\ialign{##\crcr\hfil${\scriptscriptstyle(}\mkern -1mu
    \vrule height 1.2pt width 3pt depth -.8pt {\scriptscriptstyle)}$
    \hfil\crcr \noalign{\kern-1pt\nointerlineskip}$\hfil\displaystyle{#1}\hfil$\crcr}}}
\input{babarsym}
\input{papersym}

\input{mixed_moments_num}

\long\def\inst#1{\par\nobreak\kern 4pt\nobreak
    {\it #1}\par\vskip 10pt plus 3pt minus 3pt}

\begin{document}


\begin{flushleft}
\babar-\BaBarType-\BaBarYear/\BaBarNumber \\
SLAC-PUB-\SLACPubNumber 
\end{flushleft}


\title{\large \bf
\boldmath
Measurement of Moments of the Hadronic-Mass and -Energy Spectrum in Inclusive
Semileptonic $\semilepXc$ Decays
} 

\input pubboard/authors_jun2007

\date{\today}

\begin{abstract}
We present a measurement of moments of the inclusive hadronic-mass and -energy spectrum
in semileptonic $\semilepXc$ decays. This study is based on a sample of $232$ million
$\FourS \to \BB$ decays recorded by the $\babar$ detector at the \pep2\ \epem -storage
rings. 
We reconstruct the semileptonic decay by identifying a lepton in events tagged
by a fully reconstructed hadronic decay of the second $\B$ meson. 
We report preliminary results for the moments $\mxmom{k}$
with $k=1,\ldots,6$ and $\moment{\nxn}$ with  $k=2,4,6$ and
$\nx = m_X^2 c^4 - 2 \tilde{\Lambda} E_X + \tilde{\Lambda}^2$, with $\mx$ the mass of the hadronic
system, $E_X$ its energy, and $\tilde\Lambda$ a
constant of $0.65\,\gev$, for different minimal lepton momenta between $0.8$ and $1.9 \gevc$
measured in the $\B$-meson rest frame.
These are predicted in the framework of a Heavy Quark Expansion (HQE), which allows the 
extraction of the total semileptonic branching fraction, the CKM-matrix element $\Vcb$, and
the quark masses $\mb$ and $\mc$, together with the dominant non-perturbative HQE parameters.
We find as preliminary results $\Vcb = (41.88 \pm 0.81) \cdot 10^{-3}$ and $\mb = (4.552 \pm 0.055) \gevcc$.
\\ \\
\textit{Submitted to the 2007 Europhysics Conference on High Energy Physics, 
Manchester, England.}
\end{abstract}

\pacs{12.15.Ff, 12.15.Hh, 13.25.Hw, 13.30.Ce}

\maketitle

\setcounter{footnote}{0}

\input{introduction}

\input{detector}

\input{semilep_reco}

\input{mass_moments}

\input{mixed_moments}

\input{vcb_measurement}

\input{summary}

\input{pubboard/acknowledgements}

\clearpage

\input{paperfix.bbl}
\clearpage
\input{appendix}

\end{document}

%% file: papersym.tex
\newcommand{\resultfitchisq}{8}

\newcommand{\gevsq}{\ensuremath{\mathrm{\,Ge\kern -0.1em V^2}}\xspace}
\newcommand{\gevcube}{\ensuremath{\mathrm{\,Ge\kern -0.1em V^3}}\xspace}
\newcommand{\gevtothe}[1]{\ensuremath{\mathrm{\,Ge\kern -0.1em V^{#1}}}\xspace}

\def\X       {\ensuremath{X}\xspace}
\def\Xc      {\ensuremath{X_{c}}\xspace}

\def\Breco   {\ensuremath{\B_\mathrm{reco}}\xspace}

\def\ellm    {\ensuremath{\ell^{-}}\xspace}

\def\porantip    {\kern 0.18em\optbar{\kern -0.40em p}{}\xspace}
\def\D       {\ensuremath{D}\xspace}
\def\DorDstar    {\ensuremath{\D^{\left(*\right)}}\xspace}

\newcommand{\semilepe} { \ensuremath{\Bbar \rightarrow \X e \nub}}
\newcommand{\semilepmu} { \ensuremath{\Bbar \rightarrow \X \mu \nub }}

\newcommand{\semilepXu} { \ensuremath{\Bbar \rightarrow \X_{u} \ellm \nub }}
\newcommand{\semilepXc} { \ensuremath{\Bbar \rightarrow \X_{c} \ellm \nub }}

\newcommand{\semilepD} { \ensuremath{\Bbar \rightarrow \D \ellm \nu }}
\newcommand{\semilepDstar} { \ensuremath{\Bbar \rightarrow \Dstar \ellm \nu }}
\newcommand{\semilepDstarstar} { \ensuremath{\B \rightarrow \Dstarstar \ell \nu }}
\newcommand{\semilepNreso} { \ensuremath{\B \rightarrow \DorDstar \pi \ell \nu }}

\newcommand{\BtoXsGamma} { \ensuremath{\B \rightarrow \X_{s} \g }}
\newcommand{\Gammasl}    {\ensuremath{\Gamma_{\mathit{SL}}}\xspace}

\newcommand{\Dstarstar}{\ensuremath{D^{**}}\xspace}

\def\photos      { \texttt{PHOTOS}\xspace}
\def\geantf      {\mbox{\tt GEANT4}\xspace}

\newcommand{\mxmom}[1]  { \ensuremath{\langle m_{\X}^{#1} \rangle} }

\newcommand{\mxmomreco}[1]  {\ensuremath{\langle m_{\X,reco}^{#1}\rangle }}
\newcommand{\mxmomcalib}[1]  {\ensuremath{\langle m_{\X,calib}^{#1}\rangle }}

\newcommand{\mxmomtruecut}[1]  {\ensuremath{\langle m_{\X,true}^{#1}\rangle }}
\newcommand{\mx}  {\ensuremath{m_{\X}\xspace}}
\newcommand{\mxtrue}  {\ensuremath{m_{\X,true}}}

\newcommand{\mxtruecut}  {\ensuremath{m_{\X,true,cut}}}
\newcommand{\mxreco}  {\ensuremath{m_{\X,reco}}}
\newcommand{\mxrecoi} {\ensuremath{m_{\X,reco,i}}}

\newcommand{\mxcalibi}  {\ensuremath{m_{\X,calib,i}}}

\newcommand{\elmom}[1]  { \ensuremath{\langle E_{\ell}^{#1}\rangle } }
\newcommand{\egammamom}[1]  { \ensuremath{\langle E_{\gamma}^{#1}\rangle } }

\newcommand{\Ex}  {\ensuremath{E_{\X}\xspace}}
\newcommand{\nx}  {\ensuremath{n_{\X}^{2}\xspace}}
\newcommand{\nxfour}  {\ensuremath{n_{\X}^{4}}}
\newcommand{\nxsix}  {\ensuremath{n_{\X}^{6}}}
\newcommand{\moment}[1]{\ensuremath{\langle {#1}\rangle }}
\newcommand{\mr}{\ensuremath{\mathrm}}
\newcommand{\centraltwo}{\ensuremath{ \moment{( \nx - \moment{\nx})^2}}}
\newcommand{\centralthree}{\ensuremath{ \moment{( \nx - \moment{\nx})^3}}}
\newcommand{\centraln}{\ensuremath{ \moment{( \nx - \moment{\nx})^k}}}
\newcommand{\nxn}{\ensuremath{n_{\X}^{k}}}
\newcommand{\modcentraln}{\ensuremath{ \moment{( \nx - 1.35\,\gevsq)^k}}}
\newcommand{\modcentraltwo}{\ensuremath{ \moment{( \nx - 1.35\,\gevsq)^2}}}
\newcommand{\modcentralthree}{\ensuremath{ \moment{( \nx - 1.35\,\gevsq)^3}}}
\newcommand{\centralctwo}{\ensuremath{ \moment{( \nx - C)^2}}}
\newcommand{\centralcthree}{\ensuremath{ \moment{( \nx - C)^3}}}

\newcommand{\Pmissfourmom}     { \ensuremath{ P_\mathrm{miss} } }

\newcommand{\epmiss}     { \ensuremath{E_\mathrm{miss} - c |\vec{p}_\mathrm{miss}|} }
\newcommand{\epmissabs}  { \ensuremath{\left| \epmiss \right|} }
\newcommand{\emiss}      { \ensuremath{E_\mathrm{miss}} }
\newcommand{\pmiss}      { \ensuremath{|\vec{p}_\mathrm{miss}|} }

\newcommand{\plep}         { \ensuremath{p_{\ell}^{*}} }
\newcommand{\plmin}      { \ensuremath{p_{\ell,\mathrm{min}}^{*}} }
\newcommand{\plgeq}[1]   { \ensuremath{\plep \geq {#1} \gevc }}
\newcommand{\plbin}[2]   { \ensuremath{ {#1} \leq \plep <  {#2} \gevc }}
\newcommand{\plcut}      { \ensuremath{p_{\ell,\mathrm{min}}^{*}} }

\newcommand{\El}      { \ensuremath{E_{\ell}^{*}} }

\newcommand{\Egammacut}      { \ensuremath{E_{\gamma,\mathrm{min}}} }
\newcommand{\Egammageq}[1]      { \ensuremath{E_{\gamma} \geq {#1} \gev} }

\newcommand{\brf}        { \ensuremath{\mathcal{B}}}

\newcommand{\MultX}      { \ensuremath{N_{\Xc}} }

\def\mesmax     {\mbox{$m_{\mathrm{ES,max}}$}\xspace}

\newcommand{\Ctrue} { \ensuremath{\mathcal{C}_{true}}}
\newcommand{\Ccalib} { \ensuremath{\mathcal{C}_{calib}}}

\def\wi         { \ensuremath{w_{i}}\xspace }

\def\cov        { \ensuremath{\mathcal{C}}}

\newcommand{\mb}         { \ensuremath{m_{\b}}\xspace }
\newcommand{\mc}         { \ensuremath{m_{\c}}\xspace }

\newcommand{\mupi}      { \ensuremath{\mu_{\pi}^{2}}\xspace }
\newcommand{\muG}      { \ensuremath{\mu_{G}^{2}}\xspace }
\newcommand{\rhoD}      { \ensuremath{\rho_{D}^{3}}\xspace }
\newcommand{\rhoLS}      { \ensuremath{\rho_{\mathit{LS}}^{3}}\xspace }

\newcommand{\covhqe}{ \ensuremath{\cov_{\mathrm{HQE}}}} 
\newcommand{\covexp}{ \ensuremath{\cov_{\mathrm{exp}}}} 
\newcommand{\covtot}{ \ensuremath{\cov_{\mathrm{tot}}}}

%% file: mixed_moments_num.tex
\newcommand{\epmisslow}{0}
\newcommand{\epmissup}{0.3}
\newcommand{\emisslow}{0}
\newcommand{\pmisslow}{0}
\newcommand{\percentResidualBG}{10\,\%\xspace}
\newcommand{\percentCombBG}{12\,\%\xspace}
\newcommand{\percentBG}{22\,\%\xspace}
\newcommand{\NsigFirst}{$7827 \pm 108$}
\newcommand{\NsigSecond}{$1278 \pm 42$}
\newcommand{\resolNx}{1.31\,\gevsq}
\newcommand{\biasNx}{-0.08\,\gevsq}

%% file: pubboard/authors_jun2007.tex
%
\author{B.~Aubert}
\author{M.~Bona}
\author{D.~Boutigny}
\author{Y.~Karyotakis}
\author{J.~P.~Lees}
\author{V.~Poireau}
\author{X.~Prudent}
\author{V.~Tisserand}
\author{A.~Zghiche}
\affiliation{Laboratoire de Physique des Particules, IN2P3/CNRS et Universit\'e de Savoie, F-74941 Annecy-Le-Vieux, France }
\author{J.~Garra~Tico}
\author{E.~Grauges}
\affiliation{Universitat de Barcelona, Facultat de Fisica, Departament ECM, E-08028 Barcelona, Spain }
% ...
\collaboration{The \babar\ Collaboration}
\noaffiliation

%% file: introduction.tex
\section{Introduction}
\label{sec:introduction}

Measurement of moments of the hadronic-mass
\cite{Csorna:2004CLEOMoments, 
Aubert:2004BABARMoments, 
Acosta:2005CDFMoments, 
Abdallah:2005DELPHIMoments, 
Schwanda:2007BELLEMassMoments}
and lepton-energy
\cite{Aubert:2004BABARLeptonMoments, 
Abdallah:2005DELPHIMoments,
Abe:2005BELLELeptonMoments}
spectra in inclusive semileptonic decays \semilepXc\ have been
used to determine the non-perturbative QCD parameters describing these decays
and the CKM matrix element \Vcb.

Combined fits to these moments and moments of the photon-energy spectrum
in $\BtoXsGamma$ decays \cite{Chen:2001CLEOXsGammaMoments, 
Koppenburg:2004BELLEXsGammaInclusive, Aubert:2005BABARXsGammaExclusive,
Aubert:2006XsGammaInclusive}
in the context of Heavy Quark Expansions (HQE) of QCD have resulted in
precise determinations of $\Vcb$ and $\mb$, the mass of the
$\b$ quark. Specifically, they are reported to be
$\Vcb = (42.0\pm 0.2\pm 0.7) \cdot 10^{-3}$ and  $\mb = (4.59 \pm 0.03 \pm 0.03) \gevcc$
in the kinetic mass scheme \cite{Buchmuller:2005globalhqefit} and
$\Vcb = (41.4 \pm 0.6 \pm 0.1) \cdot 10^{-3}$ and $\mb = (4.68 \pm 0.03) \gevcc$
in the 1S scheme \cite{Bauer:2004GlobalFit1SScheme}.

Lepton-energy moments are known with very good accuracy, but the precision 
of the hadronic-mass and photon-energy moments is limited by
statistics. Therefore, we present here an updated measurement of the
hadronic-mass moments $\mxmom{k}$ with $k=1,\ldots,6$ based on a
larger dataset than previously used \cite{Aubert:2004BABARMoments}.
In addition we present measurements of the mixed hadron mass-energy moments
$\moment{\nxn}$ with $k=2,4,6$  as proposed by Gambino and Uraltsev 
\cite{Gambino:2004MomentsKineticScheme}. All moments are presented
for different cuts on the minimum energy of the charged lepton.
The mixed moments use the mass \mx\ and the energy \Ex\ of the \Xc\ system in the
\B meson rest frame of \semilepXc\ decays,
\begin{equation}\label{eq:nxDef}
    n_X^2 = m_X^2 c^4 - 2 \tilde{\Lambda} E_X + \tilde{\Lambda}^2,
\end{equation}
with  a constant $\tilde\Lambda$, here fixed to be 0.65\,\gev as proposed in \cite{Gambino:2004MomentsKineticScheme}.
They allow a more reliable extraction of the higher-order non-perturbative HQE
parameters and thus they are expected to increase the precision on the extraction of $\Vcb$ and the
quark masses $\mb$ and $\mc$.

We perform a combined fit to the hadronic mass moments, measured moments of
the lepton-energy spectrum, and moments of the photon energy spectrum in
decays $\BtoXsGamma$. The fit extracts values for $\Vcb$, the quark masses $\mb$ and $\mc$,
the total semileptonic branching fraction $\brf(\semilepXc)$, and the dominant
non-perturbative HQE parameters. These are $\mupi$ and $\muG$,
parameterizing effects at ${\cal O}(1/\mb^2)$, and $\rhoD$ and $\rhoLS$,
parameterizing effects at ${\cal O}(1/\mb^3)$.

%% file: detector.tex
\section{\babar\ Detector and Dataset}
\label{sec:detector}

The analysis is based on data collected with the \babar\ experiment \cite{Aubert:2001detector}
at the \pep2 asymmetric-energy \epem storage rings \cite{Slac:1993pep2} at the
Stanford Linear Accelerator Center between October 1999 and July 2004. 

The \babar\ tracking system used for charged particle and vertex
reconstruction has two main components: a silicon vertex tracker
(SVT) and a drift chamber (DCH), both operating within a 1.5-T
magnetic field of a superconducting solenoid.
The transverse momentum resolution is 0.47\,\% at 1\gevc.
Photons are identified in an electromagnetic calorimeter (EMC)
surrounding a detector of internally reflected Cherenkov light
(DIRC), which associates Cherenkov photons with tracks for particle
identification (PID). 
The energy of photons is measured with a resolution of 3\,\% at 1\gev.
Muon candidates are identified with the
use of the instrumented flux return (IFR) of the solenoid.
The detector covers the polar angle of $30^\circ < \theta < 140^\circ$ in the center of mass (c.m.) frame. 

The data sample consists of about 210.4\,\invfb, corresponding
to $(232 \pm 3) \times 10^{6}$  decays of \FourS $\to$ \BB.
We use Monte Carlo (MC) simulated events to determine
background distributions and to correct for detector acceptance and
resolution effects. The simulation of the $\babar$ detector is realized 
with $\geantf$ \cite{Agostinelli:2002Geant4}. 
Simulated \B\ meson decays are generated using \evtgen \cite{Lange:2001EvtGen}.
Final state radiation is modeled
with $\photos$ \cite{Richter-Was:1992PHOTOS}.

The simulations of $\semilepXc$ decays use a parameterization of 
form factors for 
$\Bb\to D^{*}\ell^-\nub$~\cite{Duboscq:1996FormFactor}, 
and models for 
$\Bb\to D \ell^-\nub,D^{**}\ell^-\nub$~\cite{Scora:1995FormFactor} and
$\Bb\to D \pi \ell^-\nub, D^* \pi \ell^-\nub$~\cite{Goity:1995SoftPion}.


%% file: semilep_reco.tex
\section{Reconstruction of Semileptonic Decays}
\label{sec:semilep_decays}

\subsection{Selection of Hadronic $\B$-Meson Decays}
\label{sec:hadronic_mass_moments:breco}

The analysis uses \FourS\to\BB\ events in which one of the \B mesons 
decays to hadrons and is fully reconstructed ($\Breco$) 
and the semileptonic decay of the recoiling \Bb\ meson ($B_{\rm recoil}$) 
is identified by the presence of an electron or muon. While this approach
results in a low overall event selection efficiency of only a few per mille,
it allows for the determination of the momentum, charge, and flavor of the \B mesons.
To obtain a large sample of $B$ mesons,  many exclusive hadronic decays 
are reconstructed~\cite{Aubert:2003VubRecoil}. The kinematic consistency of 
these $\Breco$ candidates is checked with two variables,
the beam-energy-substituted mass 
$\mes = \sqrt{s/4 - \vec{p}^{\,2}_B}$ and the energy difference
$\Delta E = E_B - \sqrt{s}/2$. Here $\sqrt{s}$ is the total
energy in the c.m.\ frame, $\vec{p}_B$ and $E_B$
denote the c.m.\ momentum and c.m.\ energy of the $\Breco$ candidate.
We require $\Delta E = 0$ within three standard
deviations,  which range between 10 and 30\mev depending on the number of hadrons
used for the reconstruction.
For a given \Breco\ decay mode, the purity
is estimated as the signal fraction in events  with \mes$>5.27$\gevcc.
For events with one high-momentum lepton with \plgeq{0.8}\ in the $\B$-meson rest frame,
the purity is approximately 78\,\%.

\subsection{Selection of Semileptonic Decays}
\label{sec:hadronic_mass_moments:selection}

Semileptonic  decays are identified by the presence of one and 
only one electron or muon above a minimum momentum $\plmin$  
measured in the rest frame of the $B_{\rm recoil}$.
Electrons are selected 
with 94\% average efficiency and a hadron misidentification rate in the order of 0.1\%. 
Muons are identified 
with an efficiency ranging between 60\% for momenta $p = 1\gevc$ in
the laboratory frame and 75\% for momenta $p > 2\gevc$
and a hadron misidentification rate between 1\% for kaons and protons and 3\% for pions.
Efficiencies and misidentification rates are estimated
from selected samples of electrons, muons, pions, and kaons.
We impose the condition $Q_b Q_{\ell} < 0$, where $Q_{\ell}$ is 
the charge of the lepton and $Q_b$ is the charge of the  $b$-quark 
of the $B_{\rm reco}$. This condition is fulfilled for primary leptons, 
except for \BzBzb\ events in which flavor mixing has occurred.
We require the total observed charge of the event to be 
$|Q_{\rm tot}|= |Q_{\rm B_{reco}} + Q_{\rm B_{recoil}}| \leq 1$,
allowing for a charge imbalance in events with low momentum tracks 
or photon conversions. In cases where only one charged track is present in the reconstructed
\Xc\ system,  the total charge in the event is required to be equal to zero.

\subsection{Reconstruction of the Hadronic System}

The hadronic system $\Xc$ in the decay \semilepXc\ is reconstructed from charged
tracks  and energy depositions in the calorimeter that are not associated with
the $\Breco$ or the charged lepton. Procedures are implemented to eliminate 
fake tracks, low-energy beam-generated photons, 
and energy depositions in the  calorimeter originating from hadronic showers
faking the presence of additional particles.  Each track is assigned a specific
particle type, either $\porantip$, $\Kpm$, or $\pipm$, based on combined information
from the different $\babar$ subdetectors. The four-momentum of the reconstructed
hadronic system $P_{\Xc}$ is calculated from the four-momenta of the
reconstructed tracks $P_{i,trk}$,  reconstructed using the mass of the
identified particle type, and photons $P_{i,\gamma}$ by
$P_{\Xc} = \sum_{i=1}^{N_{trk}} P_{i,trk} + \sum_{i=1}^{N_{\gamma}} P_{i,\gamma}$.
The hadronic mass $\mx$ is calculated from the reconstructed four-momenta as
$\mx = \sqrt{ P_{\Xc}^{2} }$.

The four-momentum of the unmeasured neutrino $P_{\nu}$ is estimated from the
missing four-momentum  $\Pmissfourmom = P_{\FourS} - P_{\Breco} - P_{\Xc} - P_\ell$.
Here, all four-momenta are measured in the laboratory frame.
To ensure a well reconstructed hadronic system, we impose criteria on
the missing energy, $\emiss > 0.5 \gev$, the missing momentum, $\pmiss > 0.5 \gevc$,
and the difference of both quantities, $\epmissabs < 0.5 \gev$.
After having selected a well reconstructed $\Breco$ and having imposed
the selection criteria on $\epmiss$, $4.7\%$ of signal decays and $0.3\%$ of
background decays are retained.

Starting from a kinematically well defined initial state additional knowledge
of the kinematics of the semileptonic final state is used in a kinematic fit
to improve the overall resolution and reduce the bias of the measured values.
The fit imposes four-momentum conservation, the equality of the masses of the two $\B$
mesons, and constrains the mass of the neutrino, $P_{\nu}^{2} = 0$.
The resulting average resolutions in $\mx$
and $\nx$ are $0.355 \gevcc$ and $1.31 \gev^{2}$, respectively.
The overall biases of the kinematically fitted
hadronic system are found to be $-0.096 \gevcc$ and $-0.08 \gev^{2}$, respectively.
We require the fit to converge, thus ensuring that the constraints are fulfilled.

\begin{figure}
   \begin{center}
   \includegraphics{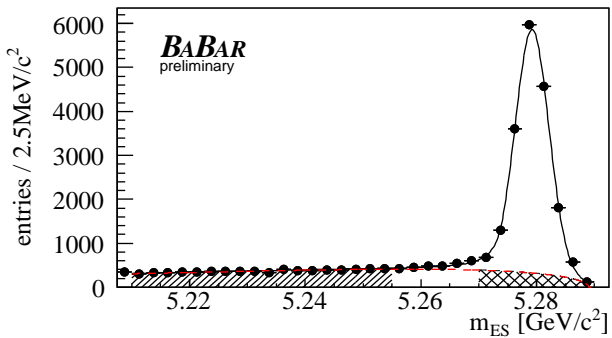}
   \end{center}
   \caption{$\mes$ spectrum of $\Breco$ decays accompanied by a lepton with $\plgeq{0.8}$.
            The signal (solid line) and background (red dashed line) components of the fit
            are overlaid. The crossed area shows the background under the $\Breco$ signal.
            The background control region in the $\mes$ sideband is indicated by
            the hatched area.
            }
    \label{fig:mesFits}
\end{figure}

The background is composed of $\epem \rightarrow \q\qbar\, (\q = u,d,s,c)$ events
(continuum background) and decays $\FourS \rightarrow \BpBm$  or $\BzBzb$ in which 
the $\Breco$ candidate is mistakenly  reconstructed from particles coming from both
$\B$ mesons in the event (combinatorial background).
Missing tracks and photons in the reconstructed hadronic system
are not considered an additional source of background since they only affect 
its resolution.
The effect of missing particles in the reconstruction is taken care 
of by further correction procedures. 
To quantify the amount of background in the $\mes$ signal region
we perform a fit to the $\mes$ distribution of the $\Breco$ candidates.
We parametrize the background using an empirical threshold
function \cite{Albrecht:1987argusFunction},
\begin{equation}
    \frac{ \text{d}N }{\text{d} \mes } \propto \mes \sqrt{ 1 - x^{2} }
           e^{ -\chi \left( 1 - x^{2} \right) },
\end{equation}
where $x = \mes / \mesmax$, $\mesmax = 5.289\gevcc$ is the kinematic endpoint approximated by the mean c.m. energy,
and $\chi$ is a free parameter defining the curvature of the function.
The signal is parameterized with a modified Gaussian function \cite{Skwarnicki:1986cbFunction}
peaked at the $\B$-meson mass and corrected for radiation losses.
The fit is performed separately for several bins
in $\mx$ and $\nx$ to account for changing background contributions
in different $\mx$ or $\nx$ regions, respectively. The background shape
is determined in a signal-free region of the $\mes$ sideband,
$5.21 \leq \mes \leq 5.255 \gevcc$. Figure \ref{fig:mesFits} shows the
$\mes$ distribution for $\plgeq{0.8}$ together with the fitted signal and
background contributions.

Residual background is estimated from MC simulations. It is composed of
charmless semileptonic decays $\semilepXu$, hadrons misidentified
as leptons, secondary leptons from semileptonic
decays of $\DorDstar$, $\Ds$ mesons or $\tau$ either in $\BzBzb$ mixed events
or produced in $\b \rightarrow \c \cbar \s$ transitions, as well as leptons from
decays of $\jpsi$, and $\psitwos$. The simulated background spectra are
normalized to the number of $\Breco$ events in data. We verify the
normalization using an independent data control sample with inverted
lepton charge correlation, $Q_b Q_{\ell} > 0$.

%% file: mass_moments.tex
\section{Hadronic Mass Moments}
\label{sec:hadronic_mass_moments}

We present measurements of the moments $\mxmom{k}$, with $k=1,\ldots6$,
of the hadronic mass distribution in semileptonic $\B$ meson decays $\semilepXc$.
The moments are measured as functions of the
lower limit on the lepton momentum, $\plmin$, between
$0.8\gevc$ and $1.9\gevc$ calculated in the rest frame of the $\B$ meson.

\subsection{Selected Event Sample}

\begin{figure}
   \begin{center}
   \includegraphics{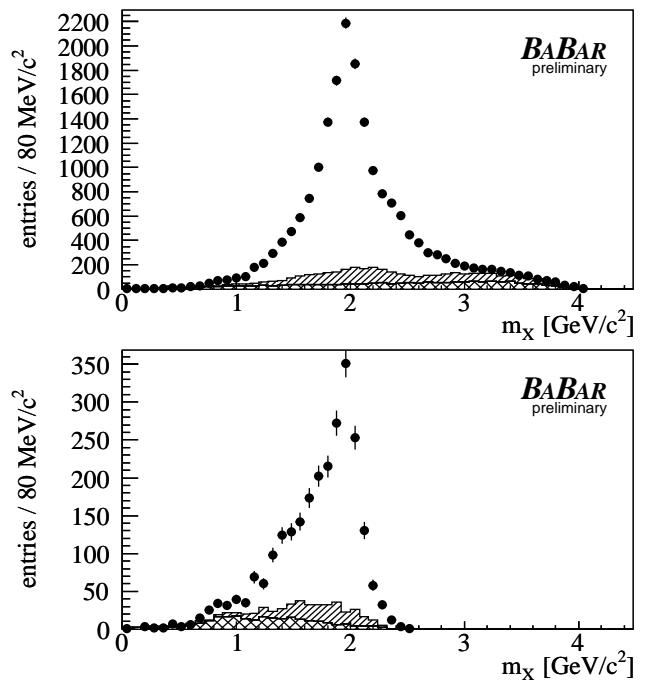}
   \end{center}
   \caption{Kinematically fitted hadronic mass spectra
            for minimal lepton momenta $\plgeq{0.8}$ (top) and $\plgeq{1.9}$ (bottom)
            together with distributions of combinatorial background
            and background from non-$\BB$ decays (hatched area) 
            as well as residual background (crossed area)
           }
    \label{fig:mass_spectra}
\end{figure}

The selected event sample contains about $21.5\%$ background.
For $\plgeq{0.8}$ we find a total of $15085 \pm 146$ signal events above
a combinatorial and continuum background of $2429 \pm 43$ events and
residual background of $1696 \pm 19$ events. For $\plgeq{1.9}$ we find
$2006 \pm 53$ signal events above a background constituted of $271 \pm 17$ and
$248 \pm 7$ combinatorial and residual events, respectively.
Figure \ref{fig:mass_spectra} shows the kinematically fitted 
$\mx$ distributions together with the extracted background shapes for 
$\plgeq{0.8}$ and $\plgeq{1.9}$.

\subsection{Extraction of Moments}

To extract unbiased moments $\mxmom{k}$, additional corrections have to be
applied to correct for remaining effects that can distort the measured $\mx$ distribution.
Contributing effects are the limited acceptance and resolution of the $\babar$
detector resulting in unmeasured particles and in misreconstructed energies and momenta of
particles. Additionally measured particles not originating from the hadronic system
and final state radiation of leptons contribute, too.
We correct the kinematically fitted $\mx^{k}$ by applying
correction factors on an event-by-event basis using the
observed linear relationship between the moments of the measured mass $\mxmomreco{k}$
and moments of the true underlying mass $\mxmomtruecut{k}$. Correction functions
are constructed from MC simulations by calculating moments $\mxmomreco{k}$ and
$\mxmomtruecut{k}$ in several bins of the true mass $\mxtrue$ and fitting the
observed dependence with a linear function.

\begin{figure*}
   \begin{center}
   \includegraphics{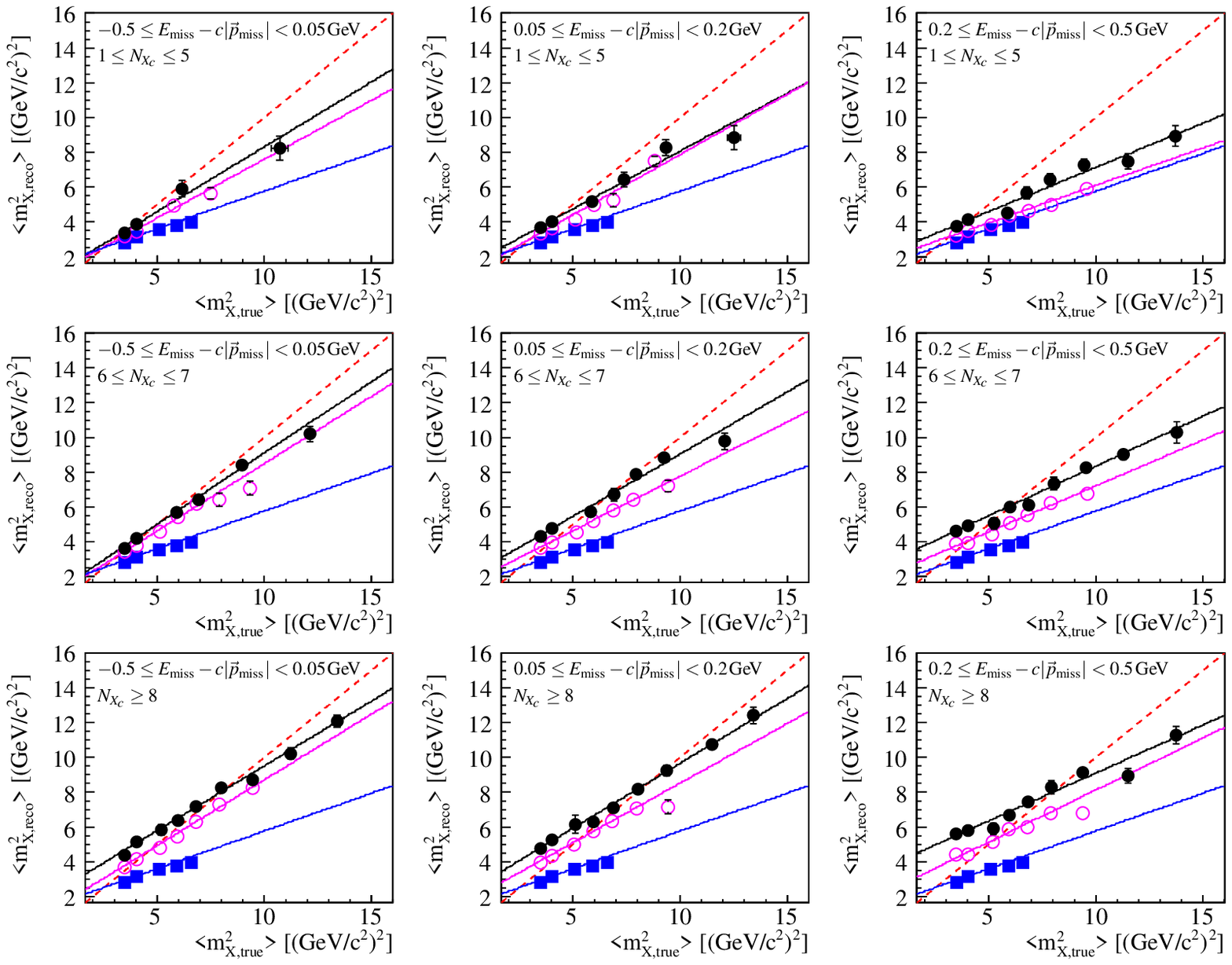}
   \end{center}
   \caption{Examples of calibration fucntions for $\mxmom{2}$ in bins of $\MultX$, $\epmiss$ and
            $\plep$. Shown are the extracted $\mxmomreco{}$ versus $\mxmomtruecut{}$ in bins
            of $\mxtrue$ for $\plbin{0.8}{0.9}$ (\textcolor{black}{$\bullet$}),
            $\plbin{1.4}{1.5}$ (\textcolor{magenta}{$\circ$}), and
            $\plgeq{1.9}$ (\textcolor{blue}{$\blacksquare$}). The results of fits of linear
            functions are overlaid as solid lines. A reference line with
            $\mxmomreco{} = \mxmomtruecut{}$ is superimposed (dashed line).
            There is only one calibration function with $\plgeq{1.9}$ constructed but
            plotted for better comparableness in each bin.
           }
    \label{fig:calib_mx2}
\end{figure*}

Studies show that the bias of the measured $\mxmomreco{k}$ is not constant over the
whole phase space but depends on the resolution and total multiplicity of the
reconstructed hadronic system, $\MultX$. Therefore, correction functions are derived
in three bins of $\MultX$, three bins of $\epmiss$, as well as in twelve bins of
$\plep$, each with a width of $100\mevc$. Due to limited number of generated MC events,
the binning in $\MultX$ and $\epmiss$ is abandoned for $\plmin \geq 1.7\gevc$.
Overall we construct $75$ calibration functions for each order of moments.
Figure \ref{fig:calib_mx2} shows examples of correction functions for the moment $\mxmom{2}$
in three bins of $\plep$ as well as in nine bins of $\epmiss$ and $\MultX$.

For each event $i$ the corrected mass $\mxcalibi^{k}$ is calculated by inverting the linear function,
\begin{equation}
    \mxcalibi^{k} = \frac{\mxrecoi^{k} - A(\epmiss, \MultX, k, \plep)}{B(\epmiss, \MultX, k, \plep)},
\end{equation}
with $A$ the offset and $B$ the slope of the correction function.
Background contributions are subtracted by applying weight factors $\wi$ dependent on
$\mxreco$ to each corrected hadronic mass, whereby each weight corresponds to
the fraction of signal events expected in the corresponding part of the
$\mxreco$ spectrum. This leads to the following expression used for the
calculation of the moments:
\begin{equation}
   \mxmom{k} = \frac{\sum\limits_{i=1}^{N_{\mathit{ev}}} \wi \, \mxcalibi^{k}}
                    {\sum\limits_{i}^{N_{\mathit{ev}}} \wi}
               \times \Ccalib \times \Ctrue.
\end{equation}
The factors $\Ccalib$ and $\Ctrue$ are dependent on the order $k$ and
minimal lepton momentum $\plmin$ of the measured moment.
They are determined in MC simulations and correct
for small biases observed after the calibration. The factors $\Ccalib$
account for the bias of the applied correction method and 
range between $0.985$ and $0.996$. For $\mxmom{6}$ we observe larger biases
ranging between $0.897$ and $0.970$ for the lowest $\plmin$ between $0.8\gevc$
and $1.2\gevc$, respectively. The residual bias correction factor $\Ctrue$
accounts for differences in selection efficiencies for different hadronic
final states and QED radiation in the final state that is included
in the measured hadron mass and distorts the measurement of the lepton's momentum.
The effect of radiative photons is estimated by employing $\photos$.
Our correction procedure results in moments which are free of photon radiation.
The residual bias correction $\Ctrue$ is estimated in MC simulations 
and typically ranges between $0.994$ and $1.014$. For the moments
$\mxmom{5}$ and $\mxmom{6}$ slighly higher correction factors are determined
ranging between $0.994$ and $1.023$ as well as $0.994$ and $1.036$, respectively.

\begin{figure}
   \begin{center}
   \includegraphics{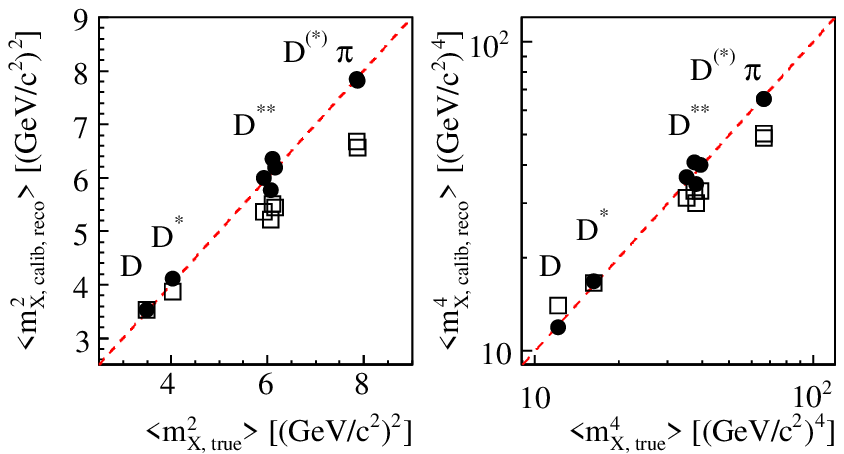}
   \end{center}
   \caption{Calibrated ($\bullet$) and uncorrected ($\Box$) moments
            $\mxmom{2}$ (left) and $\mxmom{4}$ (right)
            of individual hadronic modes for minimal lepton momenta $\plgeq{0.8}$.
            A reference line with $\mxmomcalib{} = \mxmomtruecut{}$ is superimposed.
           }
    \label{fig:massMoments_exclusiveModes}
\end{figure}

This procedure is verified on a MC sample by applying the calibration to 
measured hadron masses of individual semileptonic decays, $\semilepD$,
$\semilepDstar$, four resonant decays $\semilepDstarstar$,
and two non-resonant decays $\semilepNreso$.
Figure \ref{fig:massMoments_exclusiveModes} shows the corrected
moments $\mxmom{2}$ and $\mxmom{4}$ as functions of the true moments
for minimal lepton momenta $\plgeq{0.8}$.
The dashed line corresponds to ${\protect \mxmomcalib{k}} = {\protect \mxmomtruecut{k}}$.
The calibration reproduces the true moments over the full mass range.

\begin{figure*}
   \begin{center}
   \includegraphics{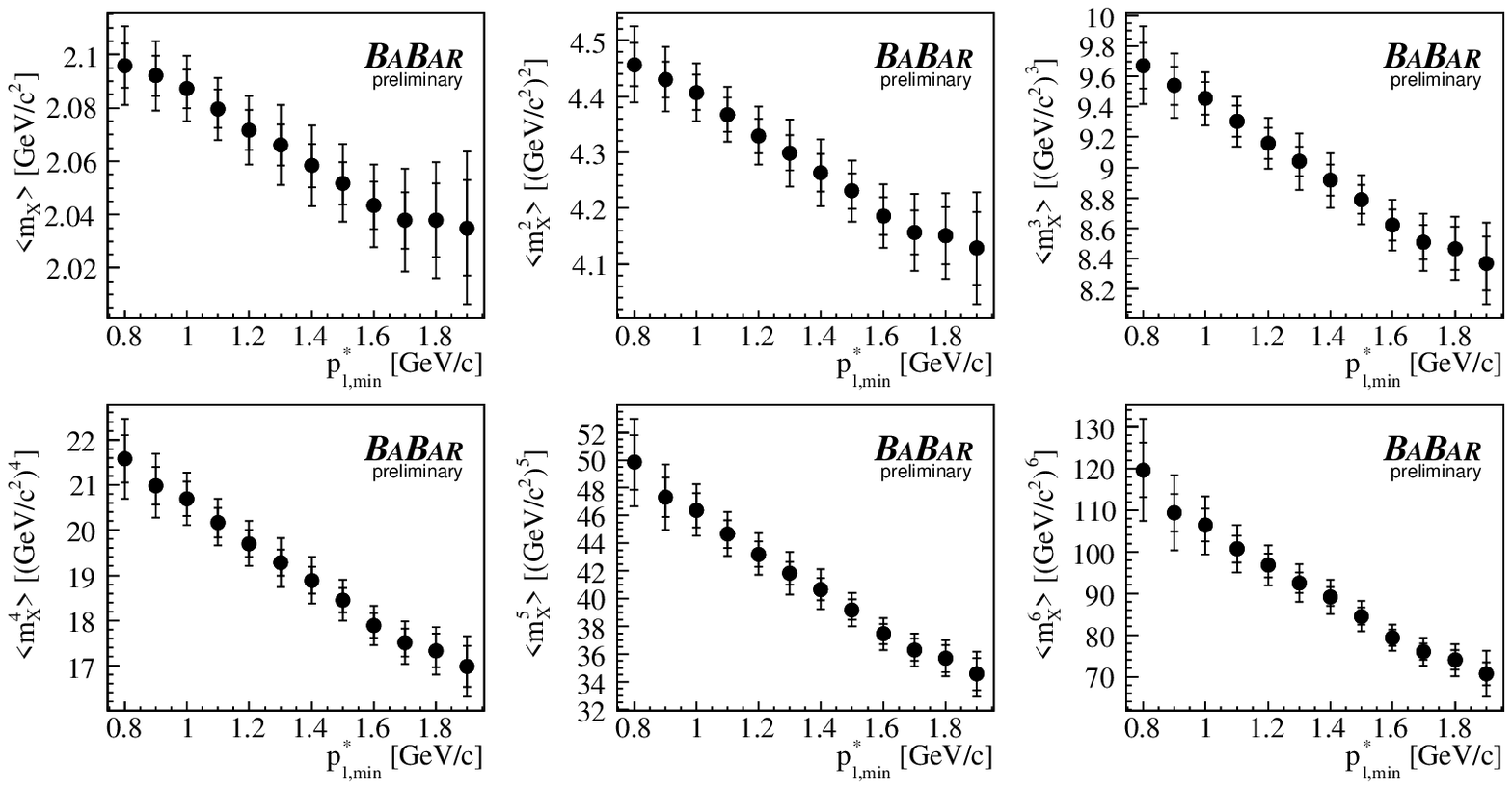}
   \end{center}
   \caption{Measured hadronic mass moments $\mxmom{k}$ with $k = 1 \ldots 6$ for different
           selection criteria on the minimal lepton momentum $\plmin$. The inner error bars
           correspond to the statistical uncertainties while the full error bars
           correspond to the total uncertainties. The moments are highly correlated.
           }
    \label{fig:massMoments}
\end{figure*}

\subsection{Systematic Studies}
\label{sec:hadronic_mass_moments:systematics}

The principal systematic uncertainties are associated with the modeling of hadronic final states
in semileptonic $\B$-meson decays, the bias of the calibration method, the subtraction
of residual background contributions, the modeling of track and photon selection efficiencies,
the identification of particles, as well as the stability of
the results. The obtained results are summarized in Tables \ref{tab:massMomentsSummary_1}
and \ref{tab:massMomentsSummary_2} for the measured moments $\mxmom{k}$ with $k = 1 \ldots 6$
and selection criteria on the minimum lepton momentum ranging from $\plgeq{0.8}$ to $\plgeq{1.9}$.

\subsubsection{Modeling of Signal Decays}

The uncertainty of the calibration method with respect to the chosen signal model
is estimated by changing the composition of the simulated inclusive hadronic spectrum.
The dependence on the simulation of high mass hadronic final states is estimated
by constructing correction functions only from MC simulated hadronic events with
hadronic masses $\mxtruecut < 2.5 \gevcc$, thereby removing the high mass tail of the simulated
hadronic mass spectrum.
The model dependence of the calibration method is
found to contribute only little to the total systematic uncertainty.
We estimate the model dependence of the residual bias correction $\Ctrue$
by changing the composition of the inclusive hadronic spectrum, thereby omitting
one or more decay modes.
We associate a systematic uncertainty to the correction of the observed bias of the
calibration method $\Ccalib$ of half the size of the applied correction.

We study the effect of differences between data and MC in the multiplicity and $\epmiss$
distributions on the calibration method by changing the binning of the
correction functions. The observed variation of the results is found to be
covered by the statistical uncertainties of the calibration functions.

\subsubsection{Background Subtraction}

The branching fractions of background decays in the MC simulation are scaled to
agree with current measurements \cite{Yao:2006pdbook}. The associated systematic
uncertainty is estimated by varying these branching fractions within their 
uncertainties. At low $\plmin$, most of the studied background channels contribute to 
the systematic uncertainty, while at high $\plmin$, the systematic uncertainty 
is dominated by background from decays $\semilepXu$.
Contributions from $\jpsi$ and $\psitwos$ decays are found to be negligible.

The uncertainty in the combinatorial $\Breco$ background subtraction is estimated by
varying the lower and upper limits of the sideband region in the $\mes$ distribution
up and down by $2.5\mevcc$. The observed effect is found to be negligible.

\subsubsection{Detector-Related Effects}

We correct the MC simulation for differences to data in the selection efficiencies of
charged tracks and photons, as well as identification efficiencies and misidentification
rates of various particle types. The corrections are extracted from data and MC
control samples.

The uncertainty of the photon selection efficiencies is found to be
$1.8\%$ per photon independent of energy, polar angle and multiplicity. The systematic
uncertainty in track finding efficiencies is estimated to be $0.8\%$ per track.
We add in quadrature the statistical uncertainty of
the control samples that depend on energy and polar angle of the track as well as the
multiplicity of tracks in the reconstructed event. The misidentification of
$\pipm$ mesons as leptons is found to affect the
overall normalization of the corresponding background spectra by $8\%$.

While the latter two uncertainties give only small contributions to the total
systematic uncertainty, the uncertainty associated with the selection efficiency
of photons is found to be the main source of systematic uncertainties.

\subsubsection{Stability of the Results}

The stability of the results is tested by dividing the data
into several independent subsamples: $\Bpm$ and $\Bz$, decays to electrons and
muons, different run periods of roughly equal data-sample sizes, and two regions
in the $\epmiss$ spectrum, $-0.5 \leq \epmiss < 0 \gev$ and $0 \leq \epmiss < 0.5 \gev$,
characterized by different resolutions of the reconstructed hadronic system.
No significant variations are observed.

The stability of the result under variation
of the selection criteria on $\epmiss$ is tested by varying the applied cut
between $\epmissabs < 0.2\gev$ and $\epmissabs < 1.4\gev$. For most of the
measured moments the observed variation is covered by other known systematic
detector and MC simulation effects. In cases where the observed variation is
not covered by those effects, we add an additional contribution to the 
systematic uncertainty of the measurement that compensates the observed difference .

\subsubsection{Simulation of Radiation}

We check the impact of low energetic photons by removing
EMC neutral energy deposits with energies below $100 \mev$ from the reconstructed hadronic
system. The effect on the measured moments is found to be negligible.

\subsection{Results}
\label{sec:hadronic_mass_moments:results}

The measured hadronic mass moments $\mxmom{k}$ with $k = 1 \ldots 6$ as functions of the
minimal lepton momentum $\plmin$ are depicted in Fig.~\ref{fig:massMoments}.
All measurements are correlated since they share subsets of selected events.
Tables \ref{tab:massMomentsSummary_1} and \ref{tab:massMomentsSummary_2} summarize the
numerical results. The statistical uncertainty consists of contributions from the data statistics
and the statistics of the MC simulation used for the construction of the correction functions,
for the subtraction of residual background, and the determination of the final bias correction.
In most cases we find systematic uncertainties that exceed the statistical uncertainty by
a factor of $~1.5$.

%% file: mixed_moments.tex
\section{Mixed Hadronic Mass- and Energy-Moments\label{sec:mixed_moments}}

The measurement of moments of the observable \nx, a combination of the mass and energy of the inclusive \Xc\ system, as defined in Eq.~\ref{eq:nxDef} , allow a more reliable extraction of the higher order HQE parameters \mupi, \muG, \rhoD, and \rhoLS. Thus a smaller uncertainty on the standard model parameters \Vcb, \mb , and \mc could be achieved.

We present measurements of the moments \moment{\nx}, \moment{\nxfour}, and \moment{\nxsix} for different minimal lepton momenta between 0.8\,\gevc and 1.9\,\gevc calculated in the \B-meson rest frame. We calculate the central moments
 \centraltwo, \centralthree, and the moments \modcentraltwo\ and \modcentralthree\ as proposed in
 \cite{Gambino:2004MomentsKineticScheme}.

Due to the structure of the variable \nx\ as a difference of two measured values, its 
measured resolution and bias are larger than for the mass moments and the
sensitivity to $\epmiss$ is increased wrt.~to $\mx$. The overall 
resolution of \nx\ after the kinematic fit for lepton momenta greater than 0.8\,\gevc 
is measured to be \resolNx\ with a bias of \biasNx. We therefore introduce stronger 
requirements on the reconstruction quality of the event. We tighten the criteria on 
the neutrino observables. The variable \epmiss is required to be between \epmisslow\ 
and \epmissup\,\gev.
Due to the stronger requirements on $\epmiss$ the individual variables
$\emiss$ and $\pmiss$ have less influence on the resolution of the
reconstrcuted hadronic system. Therefore, the cuts on the missing energy and the missing momentum in the event are loosened to $\emiss>\emisslow\,\gev$ 
and $\pmiss > \pmisslow\,\gevc$, respectively, as they do 
not yield significant improvement on the resolution of \nx, and do not increase the 
ratio of signal to background events.

The final event sample contains about \percentBG\ of background events. The background
is composed of \percentCombBG\ continuum and combinatorial background and 
\percentResidualBG\ decays of the signal \B meson other than the semileptonic
decay $\semilepXc$.
Combinatorial and continuum background is subtracted using the sideband of the \mes distribution, as described above. 
The residual background events, containing a correctly reconstructed \Breco\ meson, are subtracted using MC simulations. The dominant sources are pions misidentified  as muons, \semilepXu\ decays, and secondary semileptonic decays of $D$ and $D_s$ mesons.

\begin{figure}[t]
\centering
\includegraphics[width=0.45\textwidth]{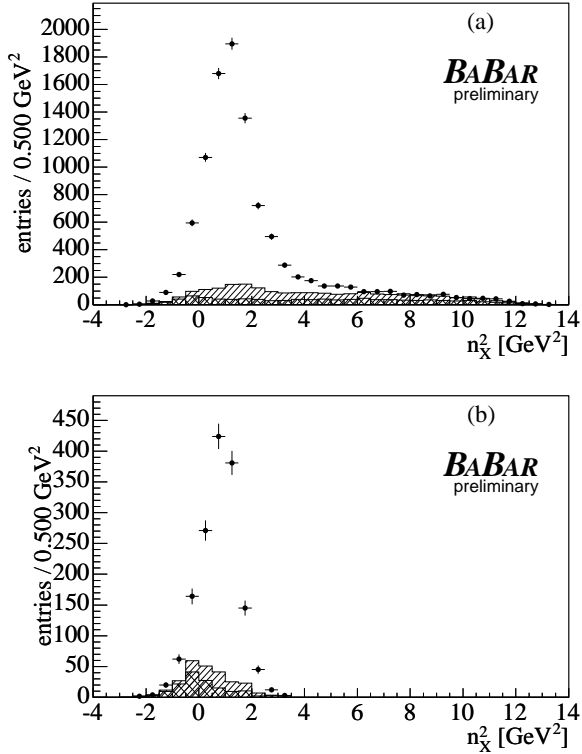}
\caption{Spectra of \nx\ after the kinematic fit together with distributions of
            combinatorial background and background from non-\BB decays
            (hatched area) as well as residual background
            (crossed area) for different minimal lepton momenta
            \plgeq{0.8} (a) and \plgeq{1.9} (b). \label{fig:nxSpectra}} 
\end{figure}

The measured \nx\ spectra for cuts on the lepton momentum  at \plgeq{0.8}\ and \plgeq{1.9} are shown together with the backgound distributions in Fig.~\ref{fig:nxSpectra}. We measure \NsigFirst\ (\NsigSecond) signal events for \plgeq{0.8\ (1.9)}, respectively.

\subsection{Extraction of Moments}

To extract unbiased moments \moment{\nxn}, effects that distort the \nx\ distribution need to be corrected. These are the limited detector acceptance, resulting in a loss of particles,
the resolution of measured charged particle momenta and energy depositions in the EMC, as well as the radiation of final-state photons. These photons are included in the measured \Xc\ system and thus lead to a modified energy and mass measurement of the inclusive system. In the case of radiation from the lepton, the lepton's measured momentum is also lowered w.r.t.\ its initial momentum.
The measured moments are corrected for the impact of these photons.

\begin{figure*}
   \begin{center}
   \includegraphics[width=0.95\textwidth]{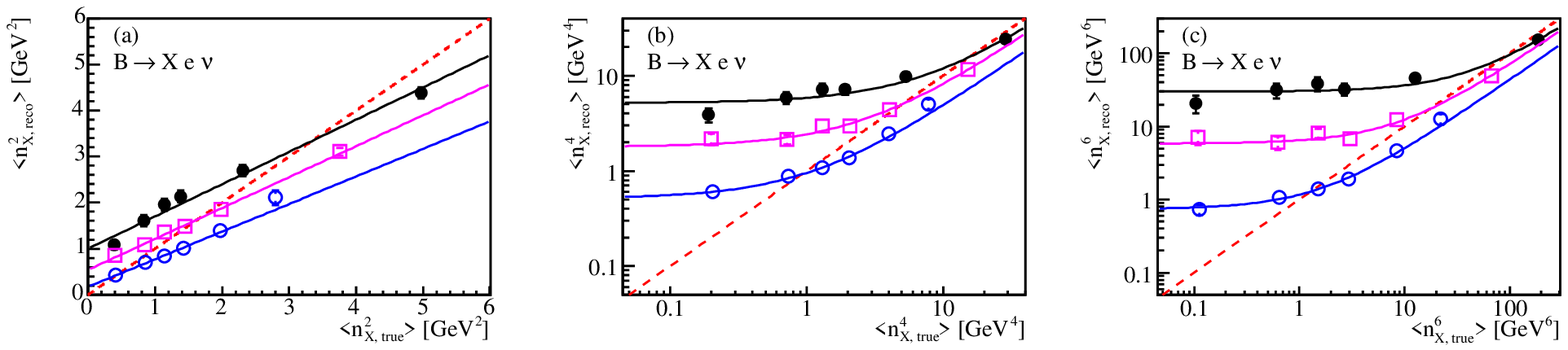}
   \includegraphics[width=0.95\textwidth]{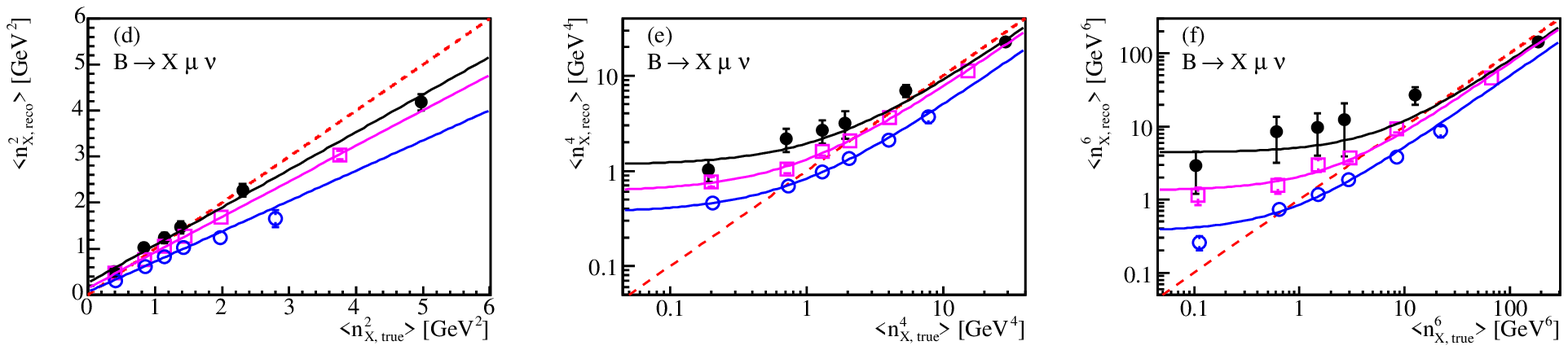}
   \end{center}
   \caption{Examples of calibration curves for \moment{\nxn} ($k=2,4,6$) in bins of  \plep, extracted separately for events \semilepe\ (a)-(c) and \semilepmu\ (d)-(f). Shown are the extracted $\moment{n^k_{X,\mathrm{reco}}}$ versus  $\moment{n^k_{X,\mathrm{true}}}$ in bins
            of ${\nx}_\mathrm{true}$ for $\plbin{0.9}{1.0}$ (\textcolor{black}{$\bullet$}),
            $\plbin{1.4}{1.5}$ (\textcolor{magenta}{$\Box$}), and
            $\plgeq{1.9}$ (\textcolor{blue}{$\circ$}) integrated over multiplicity and
            $\epmiss$ bins.
            The results of fits of linear
            functions are overlaid as solid lines. Reference lines with
            $ \moment{n^k_{X,\mathrm{reco}}} = \moment{n^k_{X,\mathrm{true}}}$
            are superimposed (dashed lines). Please note the logarithmic scales in (b), (c), (e), and (f).
            \label{fig:nxCalib}}
\end{figure*}

As described before, we find linear relationships correcting the measured means \moment{{\nxn}_\mr{reco}} to the true means \moment{{\nxn}_\mr{true}} described by first order polynomials. These functions vary with the measured lepton momentum, the measured \epmiss, and  the measured multiplicity of the inclusive \Xc system. The curves are therefore derived in three bins of \epmiss and three bins of the multiplicity for each of the 12 lepton momentum bins of 100\,\mevc.  We also find differences for events containing an electron or a muon and therefore derive separate correction functions for these two classes of events. The measured \nxn\ value is corrected on an event-by-event basis using the inverse of these functions:
\begin{equation}
n^{k}_{X,\mathrm{calib}}  = \frac{ n^{k}_{X,\mathrm{reco}} -A(\epmiss, \MultX, k, \plep)}{B(\epmiss, \MultX, k, \plep)}.
\end{equation}
Here $A$ and $B$ are the offset and the slope of the calibration function and differ for each order $k=2,4,6$ and for each of the abovementioned bins.
Figure \ref{fig:nxCalib} shows calibration curves for the moments \moment{\nxn} ($k=2,4,6$), integrated over all multiplicity bins and bins in \epmiss, for three different bins of \plep. These calibration curves are extracted separately for events containing an electron or muon. Differences are mainly visible in the low momentum bin.

\begin{figure}[b]
\centering
\includegraphics[width=0.48\textwidth]{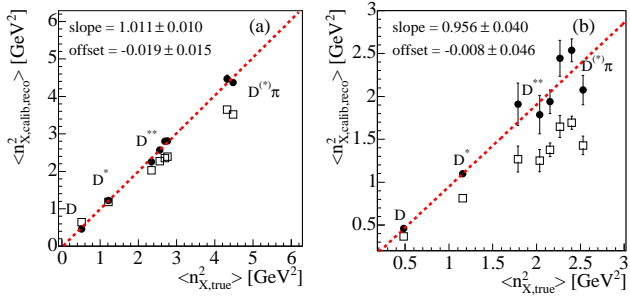}
\caption{Result of the calibration verification procedure for different minimal lepton momenta
            \plgeq{0.8} (a) and \plgeq{1.9} (b). Moments \moment{\nx}\ of exclusive modes on simulated events before calibration ($\Box$) and after calibration ($\bullet$) plotted against the true moments for each mode. The dotted line shows the fit result to the calibrated moments, the resulting parameters are shown.\label{fig:exclModes}}   
\end{figure}

To verify this calibration procedure, we extract the moments of \nxn\ of individual exclusive \semilepXc\ modes on a MC sample and compare the calibrated moments to the true moments. The result of this study for the moments 
\moment{\nx}\ is plotted in Fig.~\ref{fig:exclModes}, confirming that the extraction method is able to reproduce the true moments. Small biases remaining after calibration are of the order of 1\,\% for \moment{\nx} and in the order of few percent for \moment{\nxfour}  and \moment{\nxsix}  and are corrected and treated in the systematic uncertainties.

Background contributions are subtracted applying \nx\ dependent weight factors $w_{i}(\nx)$ on an event-by-event basis, leading to the following expression for the moments:
\begin{equation}
\moment{\nxn} = \frac{\sum\limits_{i=1}^{N_{\mathrm{ev}}} w_{i}(\nx)\cdot  {\nxn}_{\mathrm{calib},i} }     
                               {\sum\limits_{i=1}^{N_{\mathrm{ev}}} w_{i}(\nx) } 
         \times 
        \mathcal{C}(\plep, k)
\end{equation}
The bias correction factor $\mathcal{C}(\plep, k)$ depends on the minimal lepton momentum and the order of the extracted moments. It is derived on MC simulations and corrects for the small bias remaining after the calibration.

\subsection{Systematic Studies}

The main sources of systematic uncertainties have been identified as the simulation of the detector efficiency to detect neutral clusters. The corresponding effect from charged tracks is smaller but still contributes to the uncertainty on the moments. Their impact  has been evaluated by randomly excluding neutral or charged candidates from the \Xc\ system with a probability of 1.8\,\% for the neutral candidates and 0.8\% for the charged tracks, corresponding to the systematic uncertainties of the efficiency extraction methods. For the tracks we add in quadrature the statistical uncertainties from the control samples to the 0.8\,\% systematic uncertainty.
The uncertainty arising from the differences between data and MC in the $\epmiss$ distributions
is evaluated by changing the selected region of \epmiss to [0.0,0.2]\,\gev and [0.0, 0.4]\,\gev.
To evaluate the uncertainty due to the binning of the calibration curves in the multiplicity, we
randomly increase the measured multiplicity used for the choice of the calibration curve
by one with a probability of $5\%$ corresponding to observed differences between data and MC.

Smaller uncertainties arise from the unknown branching fractions of the background decay modes. Their branching fractions are scaled to agree with recent measurements \cite{Yao:2006pdbook} and are varied within their uncertainties.
The MC sample is corrected for differences in the identification efficiencies between data and MC for various particle types.
The uncertainty on the background due to pions misidentified as muons is evaluated by changing the MC corrections within the statistical uncertainties of these data control samples. While the background shape does not vary, the amount decreases up to 8\,\%. For the estimate of the uncertainty due to particle identification, we propagate this variation into the extracted moments.

A similar variation procedure is applied for the branching fractions of the exclusive signal modes, varying them several times randomly within 10\,\% for the \Dstar, 15\,\% for the $D$, 50\,\% for the individual \Dstarstar\ modes and 75\% for the non-resonant modes. The inclusive rate for the decays \semilepXc\ is conserved by rescaling all other modes. In addition, all \Dstarstar (non-resonant) modes are scaled in common, again randomly within 50\%, keeping the inclusive decay rate \semilepXc\ constant by rescaling the non-resonant (\Dstarstar) modes only.  Experimental uncertainties on the signal branching fractions are fully covered by these variations \cite{Yao:2006pdbook}. This dependence of the extraction method results in changes of the calibration curve and bias correction, however the impact on the moments measured on data is small. We conservatively add half of the bias correction remaining after calibration to the uncertainty related to the extraction method.

The stability of the results has been tested by splitting the data sample into several independent subsamples: \Bpm and \Bz, decays to electrons and muons, and different run periods of roughly equal
data-sample sizes. No significant variations are observed.
\begin{figure*}[t]
   \begin{center}
   \includegraphics[width=0.95\textwidth]{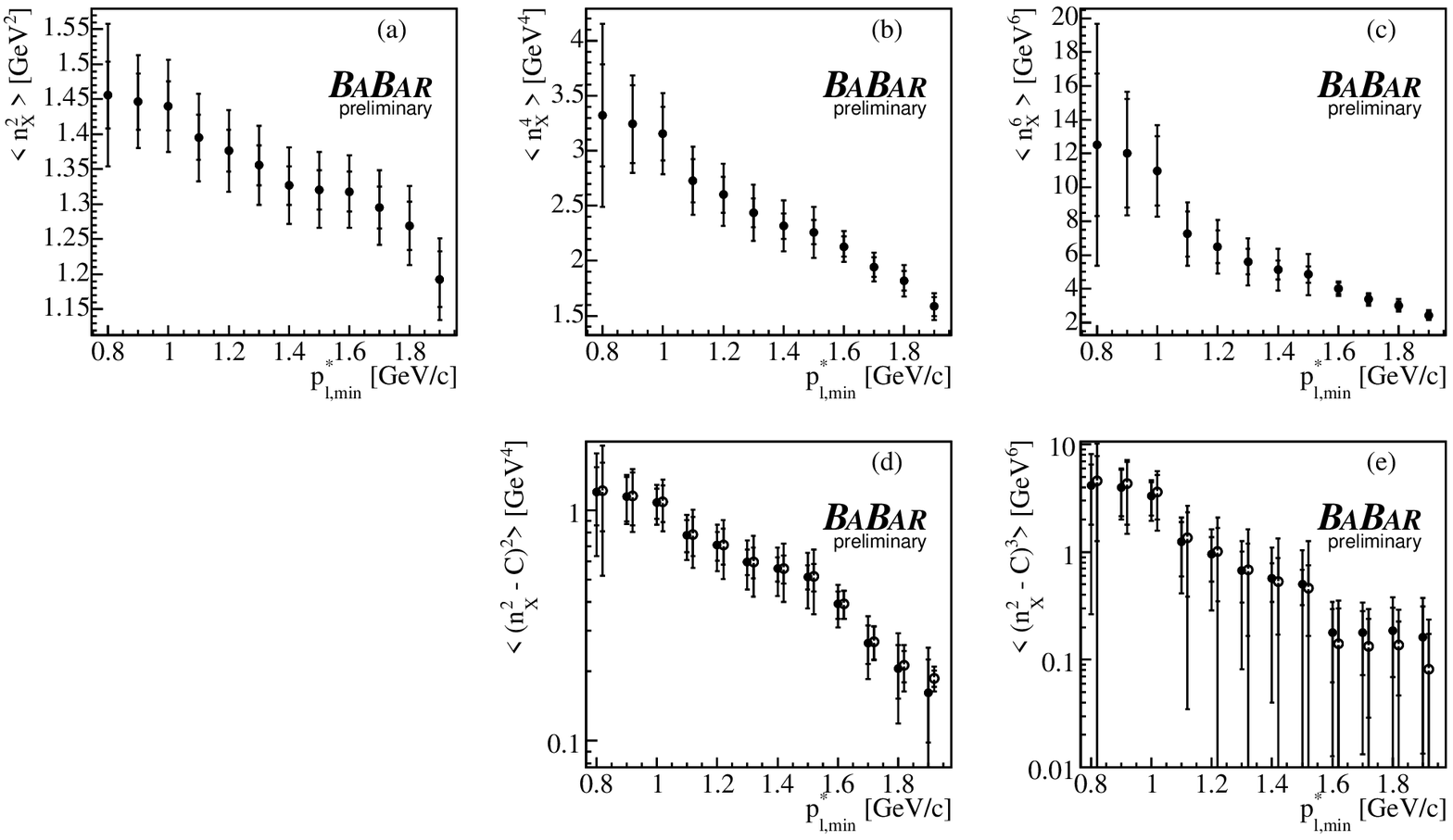}
   \end{center}
   \caption{Measured moments \moment{\nx} (a), \moment{\nxfour} (b), \moment{\nxsix} (c), 
   and the central moments \centralctwo\ with $C=\moment{\nx}$ ($\bullet$) and $C = 1.35\,\gevsq$ ($\circ$) (d),
   and \centralcthree\ with $C=\moment{\nx}$ ($\bullet$) and $C = 1.35\,\gevsq$ ($\circ$) (e) 
   for different cuts on the lepton momentum \plep. The error bars indicate the statistical and the total errors, respectively. Please note the logarithmic scale on the $y$-axis in plots (d) and (e). The moments are highly correlated. \label{fig:nxMoments}}
\end{figure*}

\subsection{Results}

Figure \ref{fig:nxMoments} shows the results for the moments  \moment{\nx}, 
\moment{\nxfour}, \moment{\nxsix}, and the central moments \centraln\ and  
\modcentraln\ for $k=2 \mathrm{\ and\ }3$ as a function of the \plep\ cut. The 
moments are highly correlated due to the overlapping
data samples. The full numerical results and the statistical and the estimated 
systematic uncertainties can be found in Tables \ref{tab:NxOrder246} -
\ref{tab:AppnxModCen3}. A clear dependence on the lepton momentum 
selection criteria is observed for all moments, due to the varying 
contributions from higher mass final states with decreasing lepton momentum.
Statistical uncertainties on the moments \moment{\nxn} arise from the limited data 
sample, the  width of the measured distribution 
$\moment{n_X^{2k}} - \moment{\nxn}^2$, and limited statistics 
on the MC samples used for the extraction of background shapes, 
calibration curves, and bias correction. In most cases we obtain 
systematic uncertainties slightly exceeding the statistical uncertainty.

%% file: vcb_measurement.tex
\section{Determination of $\Vcb$ and the quark masses $\mb$ and $\mc$}
\label{sec:vcb}

At the parton level, the weak decay rate for $\b \rightarrow \c \ell \nu$  can be calculated
accurately; it is proportional to $\Vcb^2$ and depends on the quark masses,
$m_b$ and $m_c$.  To relate measurements of the semileptonic $\B$-meson decay rate
to $\Vcb$, the parton-level calculations have to be corrected for effects of
strong interactions. Heavy-Quark Expansions (HQEs) \cite{Voloshin:1985HQE,Chay:1990HQE,Bigi:1991HQE}
have become a successful tool for calculating perturbative and non-perturbative QCD corrections
\cite{Bigi:1993PRLB293HQE,Bigi:1993PRL71HQE,Blok:1994HQE,Manohar:1994HQE,Gremm:1997HQE}
and for estimating their uncertainties.

In the kinetic-mass scheme 
\cite{Benson:2003GammaKineticScheme,Gambino:2004MomentsKineticScheme,
  Benson:2004BToSGammaKineticScheme,Aquila:2005PertCorrKineticScheme,
  Uraltsev:2004PertCorrKineticScheme,Bigi:2005KineticSchemeOpenCharm},
these expansions in $1/m_b$ and $\alpha_s(m_b)$ (the strong coupling constant)
to order ${\cal O}(1/m_b^3)$ contain six parameters: the running kinetic masses of the $b-$ and
$c-$quarks, $\mb(\mu)$ and $\mc(\mu)$, and four non-perturbative parameters.
The parameter $\mu$ denotes the Wilson normalization scale that separates
effects from long- and short-distance dynamics. The calculations are performed
for $\mu = 1 \gev$ \cite{1997:BigiKineticScheme}.
We determine these six parameters from a fit to the moments of the hadronic-mass and
electron-energy \cite{Aubert:2004BABARLeptonMoments} distributions in semileptonic $B$ decays
$\semilepXc$ and moments of the photon-energy spectrum in decays $\BtoXsGamma$
\cite{Aubert:2005BABARXsGammaExclusive,Aubert:2006XsGammaInclusive}.

In the kinetic-mass scheme the HQE to ${\cal O}(1/m_b^3)$ for the rate $\Gammasl$ of
semileptonic decays $\semilepXc$ can be expressed as \cite{Benson:2003GammaKineticScheme}

\begin{eqnarray}
   \Gammasl & = & \frac{G_F^2 \mb^5}{192\pi^3} \Vcb^2 (1+A_{\mathit{ew}}) A_{\mathit{pert}}(r,\mu)  \nonumber\\
      & \times & \Bigg [ z_0(r) \Bigg ( 1 - \frac{\mupi - \muG +
            \frac{\rhoD + \rhoLS}{c^2 \mb}}{2 c^4 \mb^2} \Bigg )  \\
      & - & 2(1-r)^4\frac{\muG + \frac{\rhoD + \rhoLS}{c^2 \mb}}{c^4 \mb^2}
        + d(r)\frac{\rhoD}{c^6 \mb^3} + \mathcal{O}(1/\mb^4)\Bigg]. \nonumber
   \label{eq:vcb_gammaslkinetic}
\end{eqnarray}

The leading non-perturbative effects arise at ${\cal O}(1/\mb^2)$ and are parameterized by
$\mupi(\mu)$ and $\muG(\mu)$, the expectation values of the kinetic and
chromomagnetic dimension-five operators. At ${\cal O}(1/\mb^3)$, two additional parameters enter,
$\rhoD(\mu)$ and $\rhoLS(\mu)$, the  expectation values of the Darwin and
spin-orbit dimension-six operators, respectively.
The ratio $r = m_c^2/m_b^2$ enters in the tree level phase-space factor
$z_0(r) = 1 - 8r + 8r^3 - r^4 - 12r^2 \ln r$ and in the function
$d(r) = 8 \ln r + 34/3 - 32r/3 - 8r^2 + 32r^3/3 - 10r^4 /3$.
The factor $1 + A_{\mathit{ew}}$ accounts for electroweak corrections. It
is estimated to be $1 + A_{ew} \cong ( 1 + \alpha/\pi \ln M_Z/\mb )^2 = 1.014$.
The quantity $A_{\mathit{pert}}$ accounts for perturbative contributions and
is estimated to be $A_{pert}(r,\mu) \approx 0.908$ \cite{Benson:2003GammaKineticScheme}.

The performed fit uses a linearized expression for the dependence
of $\Vcb$ on the values of heavy-quark parameters, expanded around
${\it a~priori}$ estimates of these parameters \cite{Benson:2003GammaKineticScheme}:
\begin{eqnarray}
        \frac{\Vcb}{0.0417}  & = & \sqrt{\frac{\brf_{clv}}{0.1032} \frac{1.55}{\tau_{\B}} } \nonumber \\
                & &\times [1 + 0.30 (\alpha_s(\mb) - 0.22) ] \nonumber\\
                & &\times [ 1 - 0.66 ( \mb - 4.60) + 0.39 ( \mc - 1.15 ) \nonumber\\
                & &+ 0.013 ( \mupi - 0.40) + 0.09 ( \rhoD - 0.20) \nonumber\\
                & &+ 0.05 ( \muG - 0.35 ) - 0.01 ( \rhoLS + 0.15 ) ].
\end{eqnarray}
Here $\mb$ and $\mc$ are in $\gevcc$ and all other parameters of the expansion are in
$\gev^{k}$; $\tau_B$ refers to the average lifetime of $\B$ mesons produced at 
the $\FourS$ and is given in \ps.
HQEs in terms of the same heavy-quark parameters are available for hadronic-mass,
electron-energy, and photon-energy moments. Predictions for those moments are obtained
from an analytical calculation. We use these calculations to determine 
$\Vcb$, the total semileptonic branching fraction $\brf$, the quark masses $\mb$ and $\mc$,
as well as the heavy-quark parameters $\mupi$, $\muG$, $\rhoD$, and $\rhoLS$,
from a simultaneous $\chisq$ fit to the measured moments and partial branching fractions,
all as functions of the minimal lepton momentum $\plmin$ and minimal photon energies 
$\Egammacut$.

\subsection{Extraction Formalism}

The fit method designed to extract the HQE parameters from the moments measurements
has been reported previously \cite{Buchmuller:2005globalhqefit, Aubert:2004BABARHQEFit}.
It is based on a $\chisq$ minimization,
\begin{equation}
    \chisq  = 
   \left( \vec{M}_{\mathrm{exp}} - \vec{M}_{\mathrm{HQE}} \right)^{T}
               \covtot^{-1}
             \left( \vec{M}_{\mathrm{exp}} - \vec{M}_{\mathrm{HQE}} \right).
    \label{eq:vcb_chi2}
\end{equation}
The vectors $\vec{M}_{\mathrm{exp}}$ and $\vec{M}_{\mathrm{HQE}}$ contain the measured
moments included in the fit and the corresponding moments calculated by theory,
respectively. Furthermore, the expression in Eq.~\ref{eq:vcb_chi2}
contains the total covariance matrix \covtot\ defined as the sum of the experimental,
\covexp, and theoretical, \covhqe, covariance matrices 
(see Section \ref{sec:vcb_theoreticalerrors}).

The total semileptonic branching fraction, $\brf(\semilepXc)$, is extracted
in the fit by extrapolating measured partial branching-fractions,
$\brf_{\plcut}(\semilepXc)$, with $\plep \geq \plmin$ to
the full lepton energy spectrum.
Using HQE predictions of the relative decay fraction
\begin{equation}
    R_{\plcut} = \frac{\int_{\plcut} \frac{\text{d}\Gammasl}{\text{d}\El} \text{d} \El}
                      {\int_{0} \frac{\text{d}\Gammasl}{\text{d}\El} \text{d} \El},
\end{equation}
the total branching fraction can be introduced as a free parameter in the fit.
It is given by
\begin{equation}
    \brf(\semilepXc) = \frac{\brf_{\plcut}(\semilepXc)}{R_{\plcut}}.
\end{equation}
The total branching fraction can be used together with the average
\B-meson lifetime $\tau_{\B}$ to calculate the total semileptonic
rate proportional to $\Vcb^{2}$,
\begin{equation}
    \Gammasl = \frac{\brf(\semilepXc)}{\tau_{\B}} \propto \Vcb^{2}.
\end{equation}
By adding $\tau_{\B}$ to the vectors of measured
and predicted quantities, $\vec{M}_{\mathrm{exp}}$ and $\vec{M}_{\mathrm{HQE}}$,
$\Vcb$ can be extracted from the fit as an additional free
parameter using Eq.~\ref{eq:vcb_gammaslkinetic}.

The non-perturbative parameters $\muG$ and $\rhoLS$ have been estimated from
$\B$-$\B^*$ mass splitting and heavy-quark sum rules to be
$\muG = (0.35 \pm 0.07) \gev^{2}$ and $\rhoLS = (-0.15 \pm 0.10) \gev^{3}$
\cite{Buchmuller:2005globalhqefit}, respectively. Both parameters are restricted
in the fit by imposing Gaussian error constraints.

\begin{figure*}
   \begin{center}
   \includegraphics{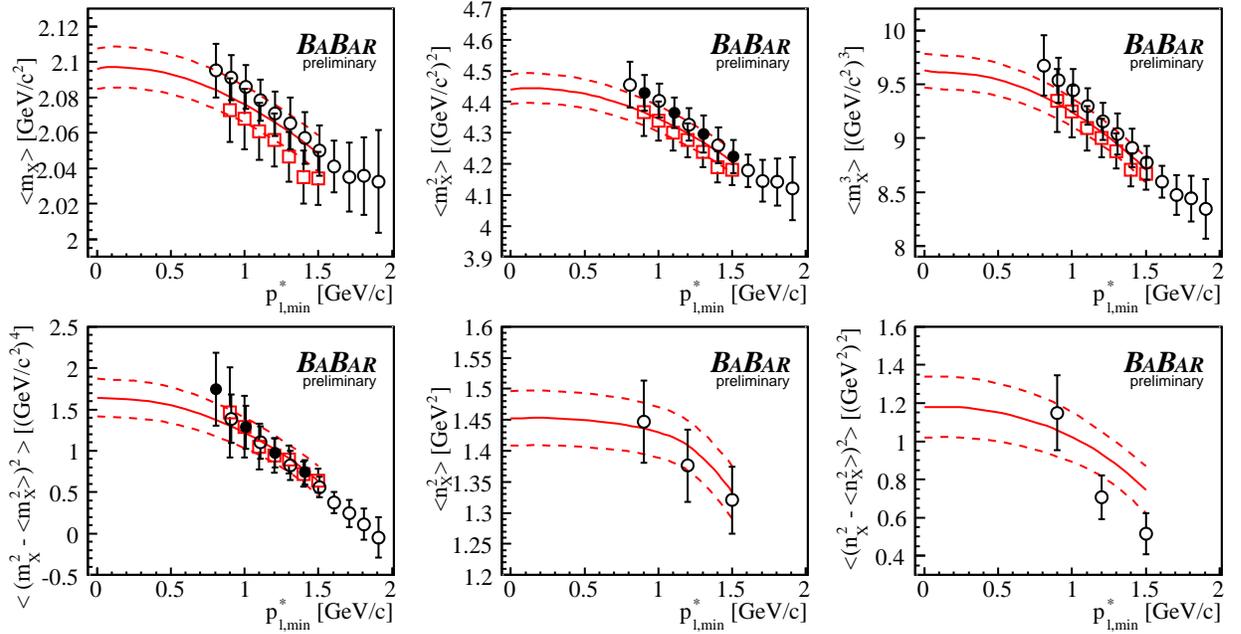}
   \end{center}
   \caption{The measured hadronic-mass and mixed moments (\textcolor{black}{$\bullet$}/\textcolor{black}{$\circ$}),
            as a function of the minimal lepton momentum $\plmin$ compared
            with the result of the simultaneous fit (solid line) and a previous measurement
            by the $\babar$ Collaboration (\textcolor{red}{$\square$}) \cite{Aubert:2004BABARMoments}.
            The solid data points mark the measurements included in the fit.
            The vertical bars indicate the experimental errors.
            The dashed lines correspond to the total fit uncertainty as
            obtained by converting the fit errors of each individual HQE parameter
            into an error for the individual moment.
           }
    \label{fig:vcb_FitMassAndMixedMoments}
\end{figure*}

\subsection{Experimental Input}

The combined fit is performed on a subset of available moment measurements
with correlations below $95\%$ to ensure the invertibility of the covariance matrix. 
Since the omitted measurements are characterized by high correlations to other
measurements considered in the fit they do not contribute significant additional 
information and the overall sensitivity of the results is not affected.
All results are based on the following set of moment measurements, 27 in total:
\begin{itemize}

 \item Lepton energy moments measured by \babar\ \cite{Aubert:2004BABARLeptonMoments}. 
            We use the partial branching fraction $\brf_{\plcut}$ measured for \plgeq{0.6,1.0,1.5} and
            the moments \elmom{} for \plgeq{0.6,0.8,1.0,1.2,1.5}.
            The lepton energy moments \elmom{2} are used at the minimal lepton momentum \plgeq{0.6,1.0,1.5} 
            and \elmom{3} at \plgeq{0.8,1.2}.
        
  \item Hadronic mass moments are used as presented in this paper. We select the following subset for
        the fit:
            \mxmom{2} for  \plgeq{0.9,1.1,1.3,1.5} and \mxmom{4} for \plgeq{0.8,1.0,1.2, 1.4}.

  \item Photon energy moments measured in \BtoXsGamma\ decays are taken from
        \cite{Aubert:2005BABARXsGammaExclusive} and \cite{Aubert:2006XsGammaInclusive}:
        \egammamom{} for the minimal photon energy \Egammageq{1.9, 2.0} and \egammamom{2} for
        \Egammageq{1.9}.

\end{itemize}
In addition we use $\tau_{\B} = f_0 \tau_0 + (1 - f_0) \tau_{\pm} = (1.585 \pm 0.007) \ps$,
taking into account the lifetimes \cite{Yao:2006pdbook} of neutral and charged $\B$ mesons,
$\tau_0$ and $\tau_{\pm}$, and their relative production rates, $f_0 = 0.491 \pm 0.007$ \cite{Yao:2006pdbook}, the fraction of $\Bz\Bzb$ pairs.

\subsection{Theoretical Uncertainties}
\label{sec:vcb_theoreticalerrors}

As discussed in \cite{Buchmuller:2005globalhqefit} and specified in
\cite{Gambino:2004MomentsKineticScheme} the following theoretical
uncertainties are taken into account:

The uncertainty related to the uncalculated perturbative corrections
          to the Wilson coefficients of non-perturbative operators
          are estimated by varying the corresponding parameters
          $\mupi$ and $\muG$ by $20\%$ and $\rhoD$ and $\rhoLS$
          by $30\%$ around their expected values.

Uncertainties for the perturbative corrections are
          estimated by varying $\alpha_{s} = 0.22$ up and down by $0.1$ for the hadronic mass moments
          and by $0.04$ for the lepton energy moments around its nominal value.

Uncertainties in the perturbative corrections of the quark masses $m_{\b}$
          and $m_{c}$ are addressed by varying both by $20\mevcc$ up and down around
          their expected values.

For the extracted value of $\Vcb$ an additional error
          of $1.4\%$ is added for the uncertainty in the expansion of the
          semileptonic rate $\Gammasl$ 
          \cite{Benson:2003GammaKineticScheme, Bigi:2005KineticSchemeOpenCharm}.        
          It accounts for remaining uncertainties in the perturbative corrections
          to the leading operator, uncalculated perturbative corrections
          to the chromomagnetic and Darwin operator, higher order power corrections, and
          possible non-perturbative effects in the operators with charm fields.
          This uncertainty is not included in the theoretical covariance matrix
          \covhqe\ but is listed separately as a theoretical uncertainty on $\Vcb$.

          For the predicted photon energy moments $\egammamom{n}$, additional 
          uncertainties are taken into account. As outlined in
          \cite{Benson:2004BToSGammaKineticScheme},
              additional uncertainties of $30\%$ of the applied bias correction
          to the photon-energy moments and half the difference in the moments
          derived from two different distribution-function ans\"atze have to be 
          considered. Both contributions are added linearly.

The theoretical covariance matrix \covhqe\ is constructed by assuming
fully correlated theoretical uncertainties for a given moment with different
lepton momentum or photon energy cutoff and assuming uncorrelated theoretical 
uncertainties for moments of different orders and types. The additonal 
uncertainties considered for the photon energy moments are assumed to be 
uncorrelated for different moments and photon energy cutoffs.

\subsection{Results}

A comparison of the fit results for the hadronic-mass and 
mixed moments with the measured moments is shown in
Fig.~\ref{fig:vcb_FitMassAndMixedMoments}.
The moments \mxmom{} and \mxmom{3} are not included in the fit
and thus provide an unbiased comparison with the fitted HQE prediction.
We find an overall good agreement, also indicated by 
$\chisq = \resultfitchisq$ for 20
degrees of freedom. The measured moments continue to decrease with
increasing $\plmin$ and extend beyond theoretical predictions 
available for $\plmin \leq 1.5 \gevc$.

Comparing the measured moments \moment{\nx}\ and \centraltwo\
with predictions resulting from the presented fit, a good agreement 
is found. The calculations used for the predictions of the mixed
moments are currently missing $\plmin$-dependent perturbative
corrections. The $\plmin$ dependence of the perturbative corrections
for those moments is however expected to be small 
\cite{Uraltsev:2004PertCorrKineticScheme}.

The fit results for the standard model and HQE parameters are summarized in Table
\ref{tab:vcb_fitResults}. We find as preliminary results $\Vcb = (41.88 \pm 0.81) \cdot 10^{-3}$ and $\mb = (4.552 \pm 0.055) \gevcc$. The results are in good agreement with earlier determinations
\cite{Buchmuller:2005globalhqefit, Bauer:2004GlobalFit1SScheme}, showing slightly increased
uncertainties due to the limited experimental input used in this fit. 

\begin{figure*}
   \begin{center}
   \includegraphics{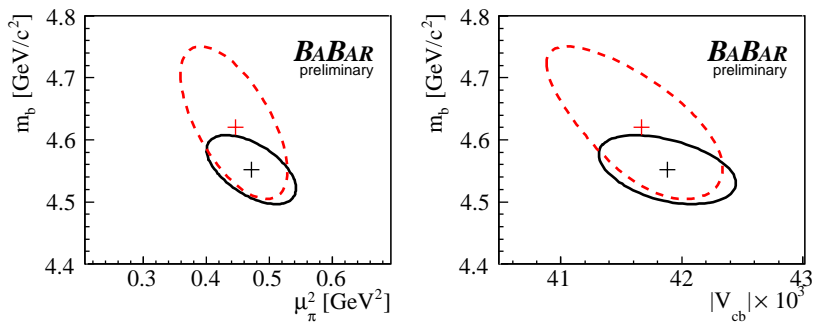}
   \end{center}
       \caption{$\Delta\chisq = 1$ contours for the fit results in the $(\mb,\Vcb)$ and $(\mb, \mupi)$
                planes comparing the results of the presented fit (black line) with those of a fit
                omitting the photon-energy moments (red dashed line).
           }
    \label{fig:vcb_contours}
\end{figure*}

Figure \ref{fig:vcb_contours} shows the $\Delta\chisq = 1$ contours in the $(\mb,\Vcb)$ and
$(\mb,\mupi)$ planes. It compares the standard fit including photon energy moments, and
a fit based on moments from semileptonic $\semilepXc$ decays only, clearly indicating the
significance of the constraints from the $\BtoXsGamma$ decays for both $\Vcb$ and $\mb$.

\begin{table*}
\caption{Fit results with experimental and theoretical uncertainties. For $\Vcb$ we take an
         additional theoretical uncertainty of $1.4\%$ from the uncertainty in the expansion
         of $\Gammasl$ into account. Correlations coefficients for all parameters are summarized
         below the results.
        }
\input{tables/fitResults.tex}
\label{tab:vcb_fitResults}
\end{table*}

%% file: tables/fitResults.tex

\begin{tabular}{lrrrrrrrr}
\hline \hline
 &$\Vcb \, \times 10^{3}$ &$\mb \, [\gevcc]$ &$\mc \, [\gevcc]$ &$\brf \, [\%]$ &$\mupi \, [\gev^{2}]$ &$\muG \, [\gev^{2}]$ &$\rhoD \, [\gev^{3}]$ &$\rhoLS \, [\gev^{3}]$  \\
\hline 
Results             &41.88 &4.552 &1.070 &10.597 &0.471 &0.330 &0.220 &-0.159  \\
$\Delta_{exp}$      &0.44  &0.038 &0.055 &0.171  &0.034 &0.042 &0.021 &0.081  \\
$\Delta_{theo}$     &0.35  &0.040 &0.065 &0.053  &0.062 &0.043 &0.042 &0.050  \\
$\Delta_{\Gammasl}$ &0.59  & & & & & & &  \\
$\Delta_{tot}$      &0.81  &0.055 &0.085 &0.179  &0.070 &0.060 &0.047 &0.095  \\
\hline 
$\Vcb$ &1.00 &-0.42 &-0.27 &0.75 &0.42 &-0.28 &0.25 &0.10  \\
$\mb$ & &1.00 &0.96 &0.09 &-0.56 &-0.07 &-0.38 &-0.24  \\
$\mc$ & & &1.00 &0.15 &-0.63 &-0.32 &-0.51 &-0.15  \\
$\brf$ & & & &1.00 &0.09 &-0.10 &0.02 &-0.04  \\
$\mupi$ & & & & &1.00 &0.40 &0.87 &0.10  \\
$\muG$ & & & & & &1.00 &0.41 &-0.05  \\
$\rhoD$ & & & & & & &1.00 &-0.21  \\
$\rhoLS$ & & & & & & & &1.00  \\
\hline 
\hline 
\end{tabular} 

%% file: summary.tex
\section{Summary}
\label{sec:summary}

We have reported preliminary results for the moments $\mxmom{k}$ with
$k = 1,\ldots,6$ of the hadronic mass distribution in semileptonic
\B-meson decays to final states containing a charm quark.
In addition we have presented preliminary results for a first measurement of the moments
$\moment{\nxn}$ for $k=2,4,6$ with $\nx$ a combination of mass and energy
of the hadronic system \Xc.
The results for the mass moments agree with the previous measurements
\cite{Csorna:2004CLEOMoments, Aubert:2004BABARMoments,
Acosta:2005CDFMoments, Abdallah:2005DELPHIMoments, Schwanda:2007BELLEMassMoments}
but tend in general to higher values, between $1\%$ and $2\%$ for $\mxmom{}$ and
$\mxmom{4}$, respectively, relative to the previous \babar\
measurement \cite{Aubert:2004BABARMoments}. The increased data sample compared to the
previous \babar\ measurement led to significantly smaller statistical
uncertainties which are smaller than the systematic uncertainties.

We have made a combined fit in the kinetic scheme to the 
hadronic mass moments, measured moments of the lepton-energy
spectrum \cite{Aubert:2004BABARLeptonMoments}, and moments of the
photon energy spectrum in decays $\BtoXsGamma$ \cite{Aubert:2005BABARXsGammaExclusive,
Aubert:2006XsGammaInclusive}.
The combined fit yields preliminary results for $\Vcb$, the quark masses $\mb$ and $\mc$,
the total semileptonic branching fraction $\brf(\semilepXc)$, and the dominant
non-perturbative HQE parameters in agreement with previous determinations. 
We obtain $\Vcb = (41.88 \pm 0.81) \cdot 10^{-3}$ and
$\mb = (4.552 \pm 0.055) \gevcc$. \\

%% file: pubboard/acknowledgements.tex
We are grateful for the 
extraordinary contributions of our \pep2\ colleagues in
achieving the excellent luminosity and machine conditions
that have made this work possible.
The success of this project also relies critically on the 
expertise and dedication of the computing organizations that 
support \babar.
The collaborating institutions wish to thank 
SLAC for its support and the kind hospitality extended to them. 
This work is supported by the
US Department of Energy
and National Science Foundation, the
Natural Sciences and Engineering Research Council (Canada),
the Commissariat \`a l'Energie Atomique and
Institut National de Physique Nucl\'eaire et de Physique des Particules
(France), the
Bundesministerium f\"ur Bildung und Forschung and
Deutsche Forschungsgemeinschaft
(Germany), the
Istituto Nazionale di Fisica Nucleare (Italy),
the Foundation for Fundamental Research on Matter (The Netherlands),
the Research Council of Norway, the
Ministry of Science and Technology of the Russian Federation, 
Ministerio de Educaci\'on y Ciencia (Spain), and the
Science and Technology Facilities Council (United Kingdom).
Individuals have received support from 
the Marie-Curie IEF program (European Union) and
the A. P. Sloan Foundation.

%% file: appendix.tex
\renewcommand{\thetable}{A.\Roman{table}}
\setcounter{table}{0}

 \begin{table*}
\addtolength{\tabcolsep}{1mm}
\caption{Results for the moments $\mxmom{k}$ with $k = 1 \ldots 3$ for different cuts on the
         minimal lepton momentum $\plep$ with absolute statistical and systematic uncertainties.
         Individual errors sources are specified due to modeling of the
         signal events, the calibration procedure, the background
         subtraction, detector efficiencies and resolution, and stability of
         moment measurements. Minimum lepton momenta cuts are given in $\gevc$.
         Moments and uncertainties are given in $(\gevcc)^{k}$.}
\input{tables/massMomentsSummary_1.tex}
\label{tab:massMomentsSummary_1}
\end{table*}

\begin{table*}
\addtolength{\tabcolsep}{1mm}
\caption{Results for the moments $\mxmom{k}$ with $k = 4 \ldots 6$ for different cuts on the
         minimal lepton momentum $\plep$ with absolute statistical and systematic uncertainties.
         Individual errors sources are specified due to modeling of the
         signal events, the calibration procedure, the background
         subtraction, detector efficiencies and resolution, and stability of
         moment measurements. Minimum lepton momenta cuts are given in $\gevc$.
         Moments and uncertainties are given in $(\gevcc)^{k}$.}
\input{tables/massMomentsSummary_2.tex}
\label{tab:massMomentsSummary_2}
\end{table*}

 

\begin{table*}
 \caption{Results for \moment{\nxn}  for $k=2,4,6$ for all cuts \plep.
 The systematic uncertainties are grouped in four categories having related sources: \textsl{rec. efficiency} is the sum of
 neutral and charged reconstruction efficiency differences data/MC, \textsl{data/MC mismod.} contains the errors from
 \epmiss differences and multiplicity differences,  \textsl{ \BR\ bg. decays} sums all contributions from the variation of
 the residual background component (including the fake lepton background), and \textsl{signal model} sums the impact of the
 variation of the signal decay branching fractions and the error related to the bias correction.
 \label{tab:NxOrder246}}
\input{tables/MixedMomentsTableDetailed_intpurity0.6_epmiss0.0-0.3_pmiss0.0_emiss0.0_trackphotonkilling_reweighted.tex}
\end{table*}


\begin{table*}
 \caption{Results for \centraltwo\ for all measured cuts on  \plep.\label{tab:AppnxCen2}}
\input{tables/CentralMomentsTableVertical_Order2.tex}
\end{table*}

\begin{table*}
 \caption{Results for \modcentraltwo\ for all measured cuts on \plep.\label{tab:AppnxModCen2}}
\input{tables/CentralMomentsModifiedTableVertical_Order2.tex}
\end{table*}

 
 \begin{table*}
 \caption{Results for \centralthree\ for all measured cuts on \plep.\label{tab:AppnxCen3}}
\input{tables/CentralMomentsTableVertical_Order3.tex} 
\end{table*}
 
\begin{table*}
 \caption{Results for \modcentralthree\ for all measured cuts on \plep.\label{tab:AppnxModCen3}}
\input{tables/CentralMomentsModifiedTableVertical_Order3.tex} 
\end{table*}

%% file: tables/massMomentsSummary_1.tex

\begin{tabular}{lrrrrrrrrr}
\hline \hline
$k$ &$p_{l,min}$ &$\langle m_{X}^{k} \rangle$ &$\sigma_{stat}$ &$\sigma_{sys}$ &Signal Model &$\mathcal{C}_{\mathrm{calib}}$ &BG subtr. &Detector &Stability  \\
\hline 
1 &0.8 &2.0958 &$\pm 0.0083$ &$\pm 0.0121$ &0.0045 &0.0042 &0.0044 &0.0095 &0.0000  \\
 &0.9 &2.0920 &$\pm 0.0075$ &$\pm 0.0107$ &0.0039 &0.0040 &0.0042 &0.0082 &0.0000  \\
 &1.0 &2.0872 &$\pm 0.0072$ &$\pm 0.0099$ &0.0038 &0.0041 &0.0041 &0.0070 &0.0009  \\
 &1.1 &2.0796 &$\pm 0.0072$ &$\pm 0.0093$ &0.0036 &0.0035 &0.0041 &0.0066 &0.0000  \\
 &1.2 &2.0717 &$\pm 0.0075$ &$\pm 0.0104$ &0.0035 &0.0047 &0.0043 &0.0067 &0.0032  \\
 &1.3 &2.0661 &$\pm 0.0078$ &$\pm 0.0128$ &0.0032 &0.0054 &0.0045 &0.0067 &0.0077  \\
 &1.4 &2.0583 &$\pm 0.0081$ &$\pm 0.0128$ &0.0028 &0.0059 &0.0048 &0.0065 &0.0075  \\
 &1.5 &2.0518 &$\pm 0.0080$ &$\pm 0.0121$ &0.0025 &0.0063 &0.0053 &0.0071 &0.0045  \\
 &1.6 &2.0433 &$\pm 0.0089$ &$\pm 0.0128$ &0.0025 &0.0077 &0.0060 &0.0079 &0.0000  \\
 &1.7 &2.0378 &$\pm 0.0105$ &$\pm 0.0162$ &0.0024 &0.0075 &0.0073 &0.0080 &0.0091  \\
 &1.8 &2.0379 &$\pm 0.0139$ &$\pm 0.0168$ &0.0025 &0.0070 &0.0089 &0.0096 &0.0075  \\
 &1.9 &2.0350 &$\pm 0.0179$ &$\pm 0.0225$ &0.0020 &0.0098 &0.0121 &0.0121 &0.0107  \\
\hline 
2 &0.8 &4.457 &$\pm 0.038$ &$\pm 0.056$ &0.022 &0.016 &0.018 &0.046 &0.000  \\
 &0.9 &4.430 &$\pm 0.032$ &$\pm 0.048$ &0.020 &0.014 &0.016 &0.038 &0.000  \\
 &1.0 &4.407 &$\pm 0.032$ &$\pm 0.041$ &0.019 &0.014 &0.015 &0.030 &0.000  \\
 &1.1 &4.368 &$\pm 0.031$ &$\pm 0.039$ &0.018 &0.011 &0.014 &0.029 &0.000  \\
 &1.2 &4.330 &$\pm 0.031$ &$\pm 0.041$ &0.017 &0.016 &0.014 &0.027 &0.015  \\
 &1.3 &4.299 &$\pm 0.032$ &$\pm 0.051$ &0.015 &0.018 &0.015 &0.027 &0.033  \\
 &1.4 &4.263 &$\pm 0.033$ &$\pm 0.049$ &0.014 &0.020 &0.016 &0.026 &0.031  \\
 &1.5 &4.231 &$\pm 0.031$ &$\pm 0.045$ &0.012 &0.021 &0.018 &0.027 &0.019  \\
 &1.6 &4.186 &$\pm 0.034$ &$\pm 0.046$ &0.012 &0.026 &0.020 &0.031 &0.000  \\
 &1.7 &4.157 &$\pm 0.040$ &$\pm 0.056$ &0.011 &0.024 &0.024 &0.031 &0.030  \\
 &1.8 &4.151 &$\pm 0.051$ &$\pm 0.058$ &0.011 &0.022 &0.029 &0.036 &0.025  \\
 &1.9 &4.128 &$\pm 0.065$ &$\pm 0.077$ &0.008 &0.035 &0.040 &0.045 &0.031  \\
\hline 
3 &0.8 &9.67 &$\pm 0.15$ &$\pm 0.21$ &0.09 &0.05 &0.06 &0.17 &0.00  \\
 &0.9 &9.54 &$\pm 0.13$ &$\pm 0.17$ &0.08 &0.04 &0.05 &0.14 &0.00  \\
 &1.0 &9.45 &$\pm 0.11$ &$\pm 0.14$ &0.07 &0.04 &0.04 &0.10 &0.00  \\
 &1.1 &9.30 &$\pm 0.10$ &$\pm 0.13$ &0.07 &0.03 &0.04 &0.10 &0.00  \\
 &1.2 &9.16 &$\pm 0.10$ &$\pm 0.13$ &0.07 &0.04 &0.04 &0.09 &0.04  \\
 &1.3 &9.04 &$\pm 0.10$ &$\pm 0.16$ &0.06 &0.05 &0.04 &0.08 &0.11  \\
 &1.4 &8.92 &$\pm 0.10$ &$\pm 0.15$ &0.05 &0.05 &0.04 &0.08 &0.10  \\
 &1.5 &8.79 &$\pm 0.09$ &$\pm 0.13$ &0.05 &0.05 &0.05 &0.08 &0.06  \\
 &1.6 &8.62 &$\pm 0.10$ &$\pm 0.13$ &0.04 &0.06 &0.05 &0.09 &0.00  \\
 &1.7 &8.51 &$\pm 0.11$ &$\pm 0.15$ &0.04 &0.06 &0.06 &0.09 &0.07  \\
 &1.8 &8.47 &$\pm 0.14$ &$\pm 0.15$ &0.04 &0.05 &0.07 &0.10 &0.05  \\
 &1.9 &8.37 &$\pm 0.18$ &$\pm 0.20$ &0.03 &0.10 &0.10 &0.13 &0.06  \\
\hline 
\end{tabular} 

%% file: tables/massMomentsSummary_2.tex

\begin{tabular}{lrrrrrrrrr}
\hline \hline
$k$ &$p_{l,min}$ &$\langle m_{X}^{k} \rangle$ &$\sigma_{stat}$ &$\sigma_{sys}$ &Signal Model &$\mathcal{C}_{\mathrm{calib}}$ &BG subtr. &Detector &Stability  \\
\hline 
4 &0.8 &21.58 &$\pm 0.52$ &$\pm 0.72$ &0.30 &0.12 &0.21 &0.61 &0.00  \\
 &0.9 &20.98 &$\pm 0.41$ &$\pm 0.58$ &0.26 &0.09 &0.15 &0.48 &0.00  \\
 &1.0 &20.69 &$\pm 0.37$ &$\pm 0.44$ &0.26 &0.09 &0.12 &0.32 &0.00  \\
 &1.1 &20.17 &$\pm 0.33$ &$\pm 0.40$ &0.24 &0.06 &0.11 &0.29 &0.00  \\
 &1.2 &19.70 &$\pm 0.30$ &$\pm 0.39$ &0.23 &0.09 &0.10 &0.26 &0.13  \\
 &1.3 &19.28 &$\pm 0.29$ &$\pm 0.46$ &0.19 &0.10 &0.10 &0.23 &0.32  \\
 &1.4 &18.89 &$\pm 0.29$ &$\pm 0.43$ &0.17 &0.11 &0.10 &0.22 &0.29  \\
 &1.5 &18.45 &$\pm 0.27$ &$\pm 0.36$ &0.15 &0.12 &0.11 &0.23 &0.18  \\
 &1.6 &17.89 &$\pm 0.27$ &$\pm 0.34$ &0.13 &0.14 &0.12 &0.25 &0.00  \\
 &1.7 &17.51 &$\pm 0.31$ &$\pm 0.37$ &0.12 &0.13 &0.14 &0.25 &0.14  \\
 &1.8 &17.33 &$\pm 0.38$ &$\pm 0.37$ &0.11 &0.11 &0.17 &0.27 &0.09  \\
 &1.9 &16.98 &$\pm 0.46$ &$\pm 0.49$ &0.08 &0.24 &0.24 &0.34 &0.04  \\
\hline 
5 &0.8 &49.83 &$\pm 1.97$ &$\pm 2.47$ &1.02 &0.40 &0.70 &2.10 &0.00  \\
 &0.9 &47.33 &$\pm 1.40$ &$\pm 1.91$ &0.90 &0.22 &0.47 &1.60 &0.00  \\
 &1.0 &46.40 &$\pm 1.23$ &$\pm 1.39$ &0.87 &0.21 &0.36 &1.00 &0.00  \\
 &1.1 &44.67 &$\pm 1.01$ &$\pm 1.23$ &0.80 &0.10 &0.29 &0.87 &0.00  \\
 &1.2 &43.22 &$\pm 0.93$ &$\pm 1.15$ &0.74 &0.19 &0.25 &0.75 &0.35  \\
 &1.3 &41.84 &$\pm 0.82$ &$\pm 1.32$ &0.62 &0.21 &0.24 &0.64 &0.92  \\
 &1.4 &40.69 &$\pm 0.82$ &$\pm 1.21$ &0.54 &0.23 &0.23 &0.59 &0.84  \\
 &1.5 &39.21 &$\pm 0.73$ &$\pm 0.99$ &0.45 &0.25 &0.25 &0.60 &0.53  \\
 &1.6 &37.46 &$\pm 0.73$ &$\pm 0.88$ &0.39 &0.30 &0.27 &0.68 &0.00  \\
 &1.7 &36.31 &$\pm 0.78$ &$\pm 0.89$ &0.35 &0.25 &0.32 &0.66 &0.26  \\
 &1.8 &35.68 &$\pm 0.96$ &$\pm 0.87$ &0.31 &0.20 &0.39 &0.68 &0.00  \\
 &1.9 &34.57 &$\pm 1.14$ &$\pm 1.17$ &0.23 &0.58 &0.53 &0.84 &0.00  \\
\hline 
6 &0.8 &119.64 &$\pm 6.56$ &$\pm 10.42$ &4.19 &6.19 &2.28 &6.89 &0.00  \\
 &0.9 &109.40 &$\pm 4.49$ &$\pm 7.80$ &3.85 &4.13 &1.43 &5.19 &0.00  \\
 &1.0 &106.41 &$\pm 3.85$ &$\pm 5.80$ &3.83 &2.86 &1.03 &3.11 &0.00  \\
 &1.1 &100.70 &$\pm 3.22$ &$\pm 4.73$ &3.34 &1.87 &0.80 &2.66 &0.00  \\
 &1.2 &96.71 &$\pm 2.75$ &$\pm 3.99$ &2.95 &1.47 &0.60 &2.17 &0.00  \\
 &1.3 &92.56 &$\pm 2.45$ &$\pm 3.77$ &2.43 &1.07 &0.55 &1.67 &2.01  \\
 &1.4 &89.20 &$\pm 2.28$ &$\pm 3.42$ &2.10 &0.65 &0.54 &1.67 &1.94  \\
 &1.5 &84.54 &$\pm 1.99$ &$\pm 3.03$ &1.84 &0.58 &0.56 &1.63 &1.58  \\
 &1.6 &79.35 &$\pm 1.90$ &$\pm 2.53$ &1.62 &0.60 &0.61 &1.75 &0.00  \\
 &1.7 &76.01 &$\pm 2.01$ &$\pm 2.58$ &1.73 &0.46 &0.71 &1.68 &0.37  \\
 &1.8 &73.97 &$\pm 2.34$ &$\pm 3.07$ &2.41 &0.35 &0.87 &1.66 &0.00  \\
 &1.9 &70.67 &$\pm 2.71$ &$\pm 4.76$ &3.91 &1.34 &1.17 &2.04 &0.00  \\
\hline 
\hline 
\end{tabular} 

%% file: tables/MixedMomentsTableDetailed_intpurity0.6_epmiss0.0-0.3_pmiss0.0_emiss0.0_trackphotonkilling_reweighted.tex

\begin{tabular}{lcrcrcrcccc}
\toprule
$k$ &\plep [\gevc] &\moment{\nxn} & &$\sigma_\mathrm{stat.}$ & &$\sigma_\mathrm{sys.}$ &rec. &data/MC &\BR &signal  \\
 & & & & & & &efficiency &mismod. & bg. decays &model  \\
\hline 
2 &0.8 &1.456 &$\pm$ &0.048 &$\pm$ &0.090 &0.054 &0.071 &0.010 &0.009  \\
 &0.9 &1.447 &$\pm$ &0.040 &$\pm$ &0.053 &0.038 &0.035 &0.010 &0.006  \\
 &1.0 &1.440 &$\pm$ &0.035 &$\pm$ &0.056 &0.042 &0.035 &0.008 &0.008  \\
 &1.1 &1.395 &$\pm$ &0.032 &$\pm$ &0.054 &0.038 &0.035 &0.006 &0.014  \\
 &1.2 &1.376 &$\pm$ &0.030 &$\pm$ &0.051 &0.035 &0.034 &0.004 &0.012  \\
 &1.3 &1.356 &$\pm$ &0.029 &$\pm$ &0.049 &0.032 &0.034 &0.004 &0.011  \\
 &1.4 &1.327 &$\pm$ &0.027 &$\pm$ &0.047 &0.032 &0.034 &0.004 &0.009  \\
 &1.5 &1.321 &$\pm$ &0.028 &$\pm$ &0.046 &0.030 &0.033 &0.004 &0.008  \\
 &1.6 &1.318 &$\pm$ &0.028 &$\pm$ &0.044 &0.027 &0.034 &0.004 &0.007  \\
 &1.7 &1.295 &$\pm$ &0.030 &$\pm$ &0.044 &0.027 &0.034 &0.005 &0.006  \\
 &1.8 &1.270 &$\pm$ &0.035 &$\pm$ &0.045 &0.028 &0.034 &0.006 &0.006  \\
 &1.9 &1.193 &$\pm$ &0.040 &$\pm$ &0.043 &0.021 &0.033 &0.007 &0.015  \\
\hline 
4 &0.8 &3.32 &$\pm$ &0.46 &$\pm$ &0.69 &0.34 &0.59 &0.08 &0.12  \\
 &0.9 &3.24 &$\pm$ &0.35 &$\pm$ &0.26 &0.17 &0.16 &0.07 &0.10  \\
 &1.0 &3.15 &$\pm$ &0.25 &$\pm$ &0.27 &0.19 &0.16 &0.05 &0.11  \\
 &1.1 &2.73 &$\pm$ &0.20 &$\pm$ &0.24 &0.13 &0.16 &0.04 &0.12  \\
 &1.2 &2.60 &$\pm$ &0.16 &$\pm$ &0.23 &0.11 &0.16 &0.01 &0.12  \\
 &1.3 &2.44 &$\pm$ &0.13 &$\pm$ &0.22 &0.09 &0.16 &0.02 &0.11  \\
 &1.4 &2.32 &$\pm$ &0.12 &$\pm$ &0.20 &0.07 &0.16 &0.01 &0.10  \\
 &1.5 &2.26 &$\pm$ &0.11 &$\pm$ &0.21 &0.08 &0.16 &0.01 &0.11  \\
 &1.6 &2.13 &$\pm$ &0.09 &$\pm$ &0.10 &0.06 &0.07 &0.00 &0.05  \\
 &1.7 &1.94 &$\pm$ &0.09 &$\pm$ &0.10 &0.05 &0.07 &0.00 &0.05  \\
 &1.8 &1.82 &$\pm$ &0.09 &$\pm$ &0.11 &0.06 &0.06 &0.00 &0.06  \\
 &1.9 &1.58 &$\pm$ &0.09 &$\pm$ &0.08 &0.05 &0.06 &0.00 &0.03  \\
\hline 
6 &0.8 &12.52 &$\pm$ &4.21 &$\pm$ &5.80 &2.83 &4.95 &0.63 &0.82  \\
 &0.9 &12.00 &$\pm$ &3.21 &$\pm$ &1.74 &1.13 &1.05 &0.55 &0.59  \\
 &1.0 &10.98 &$\pm$ &2.05 &$\pm$ &1.75 &1.22 &1.04 &0.35 &0.61  \\
 &1.1 &7.25 &$\pm$ &1.34 &$\pm$ &1.32 &0.63 &1.04 &0.22 &0.47  \\
 &1.2 &6.48 &$\pm$ &0.97 &$\pm$ &1.25 &0.54 &1.03 &0.09 &0.46  \\
 &1.3 &5.60 &$\pm$ &0.75 &$\pm$ &1.17 &0.35 &1.03 &0.08 &0.41  \\
 &1.4 &5.12 &$\pm$ &0.56 &$\pm$ &1.09 &0.17 &1.03 &0.03 &0.33  \\
 &1.5 &4.85 &$\pm$ &0.49 &$\pm$ &1.11 &0.19 &1.02 &0.02 &0.39  \\
 &1.6 &4.02 &$\pm$ &0.32 &$\pm$ &0.28 &0.16 &0.18 &0.02 &0.15  \\
 &1.7 &3.38 &$\pm$ &0.26 &$\pm$ &0.25 &0.12 &0.18 &0.01 &0.13  \\
 &1.8 &3.02 &$\pm$ &0.23 &$\pm$ &0.29 &0.14 &0.17 &0.01 &0.17  \\
 &1.9 &2.44 &$\pm$ &0.20 &$\pm$ &0.21 &0.11 &0.17 &0.02 &0.05  \\
\hline \hline
\end{tabular} 

%% file: tables/CentralMomentsTableVertical_Order2.tex

\begin{tabular}{cccc}
\toprule
\plep &$\langle (n_{X}^{2} - \langle n_{X}^{2} \rangle )^{2}\rangle$ &$\sigma_\mathrm{stat.}$ &$\sigma_\mathrm{stat.+sys.}$  \\
$[\gevc]$ &$[\gevtothe{2}]$ & &  \\
\hline 
0.8 &1.20 &0.34 &0.57  \\
0.9 &1.15 &0.25 &0.28  \\
1.0 &1.08 &0.16 &0.21  \\
1.1 &0.78 &0.12 &0.17  \\
1.2 &0.71 &0.10 &0.16  \\
1.3 &0.60 &0.08 &0.14  \\
1.4 &0.56 &0.07 &0.13  \\
1.5 &0.52 &0.06 &0.14  \\
1.6 &0.39 &0.05 &0.08  \\
1.7 &0.27 &0.05 &0.08  \\
1.8 &0.21 &0.05 &0.09  \\
1.9 &0.16 &0.06 &0.09  \\
\hline \hline
\end{tabular} 

%% file: tables/CentralMomentsModifiedTableVertical_Order2.tex

\begin{tabular}{cccc}
\toprule
\plep &$\langle (n_{X}^{2} - 1.35\,\gevsq )^{2}\rangle$ &$\sigma_\mathrm{stat.}$ &$\sigma_\mathrm{stat.+sys.}$  \\
$[\gevc]$ &$[\gevtothe{2}]$ & &  \\
\hline 
0.8 &1.21 &0.40 &0.69  \\
0.9 &1.16 &0.30 &0.35  \\
1.0 &1.09 &0.20 &0.28  \\
1.1 &0.78 &0.15 &0.22  \\
1.2 &0.71 &0.12 &0.20  \\
1.3 &0.60 &0.09 &0.18  \\
1.4 &0.56 &0.08 &0.16  \\
1.5 &0.52 &0.07 &0.16  \\
1.6 &0.39 &0.05 &0.06  \\
1.7 &0.27 &0.05 &0.04  \\
1.8 &0.21 &0.03 &0.05  \\
1.9 &0.19 &0.02 &0.02  \\
\hline \hline
\end{tabular} 

%% file: tables/CentralMomentsTableVertical_Order3.tex

\begin{tabular}{cccc}
\toprule
\plep &$\langle (n_{X}^{2} - \langle n_{X}^{2} \rangle )^{3}\rangle$ &$\sigma_\mathrm{stat.}$ &$\sigma_\mathrm{stat.+sys.}$  \\
$[\gevc]$ &$[\gevtothe{3}]$ & &  \\
\hline 
0.8 &4.19 &2.38 &3.92  \\
0.9 &3.99 &1.85 &1.99  \\
1.0 &3.33 &1.13 &1.36  \\
1.1 &1.26 &0.66 &0.84  \\
1.2 &0.96 &0.43 &0.67  \\
1.3 &0.68 &0.34 &0.59  \\
1.4 &0.57 &0.22 &0.53  \\
1.5 &0.51 &0.18 &0.53  \\
1.6 &0.18 &0.12 &0.17  \\
1.7 &0.18 &0.11 &0.16  \\
1.8 &0.19 &0.12 &0.19  \\
1.9 &0.16 &0.15 &0.21  \\
\hline \hline
\end{tabular} 

%% file: tables/CentralMomentsModifiedTableVertical_Order3.tex

\begin{tabular}{cccc}
\toprule
\plep &$\langle (n_{X}^{2} - 1.35\,\gevsq )^{3}\rangle$ &$\sigma_\mathrm{stat.}$ &$\sigma_\mathrm{stat.+sys.}$  \\
$[\gevc]$ &$[\gevtothe{3}]$ & &  \\
\hline 
0.8 &4.57 &3.31 &5.58  \\
0.9 &4.33 &2.54 &2.85  \\
1.0 &3.62 &1.59 &2.04  \\
1.1 &1.36 &0.98 &1.33  \\
1.2 &1.01 &0.66 &1.08  \\
1.3 &0.69 &0.52 &0.94  \\
1.4 &0.53 &0.36 &0.82  \\
1.5 &0.46 &0.29 &0.80  \\
1.6 &0.14 &0.16 &0.22  \\
1.7 &0.13 &0.10 &0.16  \\
1.8 &0.14 &0.09 &0.16  \\
1.9 &0.08 &0.09 &0.15  \\
\hline \hline
\end{tabular} 